\documentclass[prb,amsmath,amssymb,superscriptaddress,preprint]{revtex4}
\usepackage{float}
\usepackage{amsmath}
\usepackage{amssymb}
\usepackage{amsfonts}
\usepackage{euscript}
\usepackage{enumerate}
\usepackage{hhline}
\usepackage[para,online,flushleft]{threeparttable}
\usepackage{pslatex}
\usepackage{tabularx}
\usepackage[usenames,dvipsnames]{xcolor}
\usepackage{graphicx}
\usepackage{dcolumn}
\usepackage{bm}
\usepackage[sort&compress]{natbib}

\usepackage{lipsum}

\makeatletter
\renewcommand{\@biblabel}[1]{#1. }
\renewcommand{\@dotsep}{500}
\renewcommand{\@pnumwidth}{0em}
\renewcommand{\l@figure}[2]{
\@dottedtocline{1}{1.5em}{2em}{Figure #1}{}\vspace{15pt}}

\newcommand{\SiN}{\text{Si}_\text{3}\text{N}_\text{4}}
\newcommand{\mum}{ \,\mu\text{m}}

\usepackage[normalem]{ulem}

\begin{document}
\title{Triggered single-photon generation and resonance fluorescence in ultra-low loss integrated photonic circuits}
\author{Ashish Chanana} 
\affiliation{Microsystems and Nanotechnology Division, Physical Measurement Laboratory, National Institute of Standards and Technology, Gaithersburg, MD 20899, USA}
\affiliation{Institute for Research in Electronics and Applied Physics and Maryland NanoCenter, University of Maryland,
College Park, MD 20742, USA}
\affiliation{Theiss Research, La Jolla, CA 92037}
\author{Hugo Larocque}
\affiliation{Department of Electrical Engineering and Computer Science, Massachusetts Institute of Technology, Cambridge, MA, 02139}
\author{Renan Moreira}
\affiliation{Department of Electrical and Computer Engineering, University of California Santa Barbara, Santa Barbara, CA 93016}
\author{Jacques Carolan}
\thanks{Current address: Wolfson Institute for Biomedical Research, University College London, Gower Street, London WC1E 6BT, United Kingdom}
\affiliation{Department of Electrical Engineering and Computer Science, Massachusetts Institute of Technology, Cambridge, MA, 02139}
\author{Biswarup Guha}
\affiliation{Microsystems and Nanotechnology Division, Physical Measurement Laboratory, National Institute of Standards and Technology, Gaithersburg, MD 20899, USA}
\affiliation{Joint Quantum Institute, NIST/University of Maryland, College Park, MD 20742, USA}
\author{Vikas Anant}
\affiliation{Photon Spot, Inc.,142 W Olive Ave, Monrovia CA 91016}
\author{Jin Dong Song}
\affiliation{Center for Opto-Electronic Convergence Systems, Korea Institute of Science and Technology, Seoul 136-791, South Korea,}
\author{Dirk Englund}
\affiliation{Department of Electrical Engineering and Computer Science, Massachusetts Institute of Technology, Cambridge, MA, 02139}
\author{Daniel J. Blumenthal}
\affiliation{Department of Electrical and Computer Engineering, University of California Santa Barbara, Santa Barbara, CA 93016}
\author{Kartik Srinivasan}
\affiliation{Microsystems and Nanotechnology Division, Physical Measurement Laboratory, National Institute of Standards and Technology, Gaithersburg, MD 20899, USA}
\affiliation{Joint Quantum Institute, NIST/University of Maryland, College Park, MD 20742, USA}
\author{Marcelo Davanco}\email{marcelo.davanco@nist.gov}
\affiliation{Microsystems and Nanotechnology Division, Physical Measurement Laboratory, National Institute of Standards and Technology, Gaithersburg, MD 20899, USA}

\date{\today}

\begin{abstract}
\noindent
A central requirement for photonic quantum information processing systems lies in the combination of nonclassical light sources and low-loss, phase-stable optical modes. While substantial progress has been made separately towards ultra-low loss, $\leq1$~dB/m, chip-scale photonic circuits and high brightness single-photon sources, integration of these technologies has remained elusive. Here, we report a significant advance towards this goal, in the hybrid integration of a quantum emitter single-photon source with a wafer-scale, ultra-low loss silicon nitride photonic integrated circuit. We demonstrate triggered and pure single-photon emission directly into a $\SiN$ photonic circuit with $\approx1$~dB/m propagation loss at a wavelength of $\approx920$~nm. These losses are more than two orders of magnitude lower than reported to date for any photonic circuit with on-chip quantum emitter sources, and $>50$~\% lower than for any prior foundry-compatible integrated quantum photonic circuit, to the best of our knowledge. Using these circuits we report the observation of resonance fluorescence in the strong drive regime, a milestone towards integrated coherent control of quantum emitters. These results constitute an important step forward towards the creation of scaled chip-integrated photonic quantum information systems.
\end{abstract}  

\maketitle

\section{Introduction}
Advances have been made in photonic integrated circuit (PIC) technology based on wafer-scale ultra-low loss ($\approx1$~dB/m) waveguides (ULLWs). With propagation losses as low as 0.034 dB/m at telecommunications wavelengths~\cite{liu_720_2021}  and transparency from 405~nm through the infrared~\cite{blumenthal_photonic_2020}, the wafer-scale, CMOS compatible $\SiN$ waveguide forms the basis of a versatile and promising integration platform. While focus has been on use of such technologies for classical applications, including coherent fiber communications \cite{doerr_silicon_2018}, integrated microwave photonics \cite{marpaung_integrated_2013}, positioning and navigation \cite{lai_earth_2020} and atomic clocks \cite{newman_architecture_2019}, progress towards an ULLW integration platform for quantum applications has been limited. Overall, foundry-compatible quantum PIC platforms reported to date have featured waveguide propagation losses of $>5$~dB/m, as shown in Table ~\ref{SI_table:ULL_Q_platforms} of the Supplemental Information (SI). Low photonic losses, including both waveguide propagation and insertion losses at on-chip components such as directional couplers, are central to meeting the scaling requirements for PICs that may be used to implement practical photonic quantum simulation~\cite{sparrow}, machine learning~\cite{steinbrecher2018quantum}, and quantum computing \cite{choi2019percolation}, particularly with error correction~\cite{Rudolph:2017du}. Major loss contributions today that are detrimental to scaling include component insertion loss and waveguide interconnect loss between components like couplers, sources, and detectors. While insertion loss is a dominant factor in overall loss in quantum PICs, and must be reduced for producing throughputs comparable to those achievable in micro-optics circuits~\cite{Wang2019}, PICs with ultra-low propagation losses will likely be critical for fault-tolerant photonic computing where photons must be 'stored' in delay lines~\cite{gimeno-segovia_relative_2017}, and also for quantum simulation schemes that rely on time-demultiplexing or buffering of single-photons, such as time-bin~\cite{brod_photonic_2019} or high-dimensional Gaussian Boson Sampling~\cite{deshpande_quantum_2022}.

Bringing single-photon sources and ULLWs together on a single chip is critical for robustness,  efficiency, performance, and compactness, especially for circuits that  incorporate multiple independent sources. On-chip sources based on spontaneous four-wave mixing or spontaneous parametric down-conversion have been integrated within low-loss silicon-based and hybrid PIC platforms, with $>5$~dB/m losses (see Table~\ref{SI_table:ULL_Q_platforms} in the SI). However, these sources exhibit a fundamental trade-off between the single-photon generation probability and purity, defined as the absence of multi-photon generation events, which limits the on-chip single-photon flux ~\cite{Eisaman2011}. While multiplexing of multiple heralded sources can be employed to  overcome such trade-off~\cite{kaneda_high-efficiency_nodate}, it is challenging to simultaneously meet the phase-matching, high nonlinear coefficients and ultra-low losses with a single device layer on a chip, in particular since the requisite strong field confinement in high refractive index regions is detrimental to loss performance~\cite{ji_methods_2021}. As an alternative, single quantum emitters do not suffer from the same purity versus brightness trade-off~\cite{Somaschi2016}, and can produce pure streams of triggered single-photons at rates that are limited fundamentally by the cycling time between a ground and an excited state. Recently, integration of quantum emitter-based single-photon sources has been explored in homogeneous~\cite{Dietrich2016, Lodahl2018} or heterogeneous and hybrid PIC platforms ~\cite{elshaari_hybrid_2020, kim_hybrid_2020} with waveguide losses in excess of 1~dB/cm. New solutions are needed that bring single quantum photon emitters onto ultra-low loss, $\leq1$~dB/m, waveguide technology in a wafer-scale CMOS compatible, scalable integration platform.

\begin{figure}[h]
    \centering
    \includegraphics[width=\columnwidth]{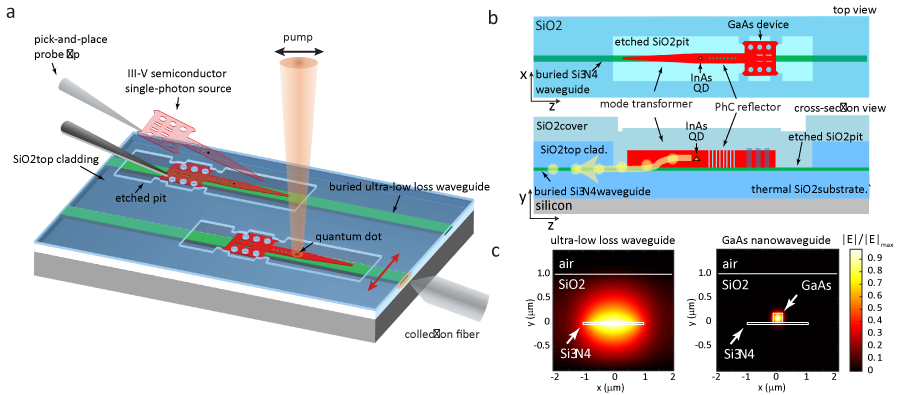}
    \caption{\textbf{a} Schematic of pick-and-place hybrid integration of a GaAs nanophotonic device containing InAs quantum dots onto an ultra-low loss $\SiN$ waveguide (ULLW). Tungsten probes were used to place and align the GaAs device to the etched pit and the buried ULLW. Control of the pump beam polarization allows resonant QD excitation with minimal pump scattering into the ULLW, allowing observation of resonance fluorescence coupled to the transverse-electric (TE) polarized mode. \textbf{b} Top-view and cross-sectional schematic of hybrid device geometry. \textbf{c} Modal profiles of the TE-polarized ULL $\SiN$ (left) and GaAs (right) waveguides } 
    \label{fig:Device} 
\end{figure}

In this work, we report a significant advance towards this goal, in demonstrating the hybrid integration of ultra-low loss PICs and quantum emitter single-photon sources. Our PICs are based on a high aspect ratio, buried channel $\SiN$  waveguide (WG) that is shown to achieve propagation losses of $\approx1$~dB/m at 920~nm, two orders of magnitude lower than reported to date for any photonic circuit with on-chip quantum emitter sources (see Table~\ref{SI_table:ULL_Q_platforms}). The quantum emitters are single InAs quantum dots (QDs) embedded in GaAs nanophotonic geometries that utilize a tapered mode-transformer to couple to $\SiN$ ultra-low loss waveguide structures~\cite{mouradian_scalable_2015, Davanco2017}. These tab-released III-V membrane structures are assembled into pockets etched in the $\SiN$ waveguide upper oxide cladding via a pick-and-place technique that has been shown to allow high-yield integration of multiple quantum emitter sources onto photonic circuits~\cite{wan_large-scale_2020}. 

To characterize our ULLWs, we utilize a photon-counting optical time-domain reflectometry (OTDR) method that allows us to determine propagation losses as low as $(1.0\pm0.40)$~dB/m along an on-chip waveguide spiral of 3~m in total length. While such a method has been employed in the past for characterizing fiber optic links~\cite{eraerds_photon_2010}, here we show that it may be used for characterizing on-chip ULLWs. 

We report the first demonstration of triggered emission of QD single-photons into ULLWs, with $g^{(2)}(0)< 0.1$, indicating high single-photon Fock-state purity. We also report the first observation of waveguide-coupled single dot resonance fluorescence in the strong drive regime, evidenced by the appearance of the Mollow triplet in the QD emission spectrum. Such a feature is a signature of dressed states emerging from the coupling of a two-level system to a strong coherent excitation field ~\cite{flagg_resonantly_2009,nick_vamivakas_spin-resolved_2009}, and is of both scientific and technological importance~\cite{ulhaq_cascaded_2012,lopez_carreno_photon_2017}.

\section{Device description and fabrication}
\label{Section:Dev_description}
Figure~\ref{fig:Device} shows a schematic of our hybrid integration platform. The ULLWs consist of a high-aspect ratio $\SiN$ core, with a thickness of 40~nm and width of 2~$\mu\text{m}$, buried under 1~$\mum$ SiO$_\text{2}$ upper cladding layer. The top cladding thickness is chosen to ensure a  weakly confined single transverse-electric (TE) guided mode with  low propagation losses in the 900~nm wavelength band~\cite{chauhan_ultra-low_2020}. The on-chip single-photon source consists of a straight GaAs nanowaveguide with embedded InAs self-assembled QDs followed by an adiabatic mode transformer, a geometry that has been shown to allow efficient coupling of QD emission directly into air-clad $\SiN$ ridge waveguides~\cite{Davanco2017,schnauber2019indistinguishable}. Opposite to the adiabatic taper, a one-dimensional photonic crystal back-reflector designed for high reflectively above 900~nm is introduced to allow unidirectional emission into the $\SiN$ waveguide. 

To ensure evanescent coupling between the GaAs and $\SiN$ layers using the mode transformer, the spacing between the transformer and waveguide  must be kept small ($<200$~nm) for air-clad $\SiN$ guides with strong confinement, as in~\cite{mouradian_scalable_2015, Davanco2017, schnauber2019indistinguishable}. Our ULLW, however, features extremely weak confinement and requires a minimum  $>250$~nm of top SiO$_2$ cladding to support a guided mode. In our approach, a pocket with special alignment features is etched into the 1~$\mum$ top SiO$_\text{2}$ upper cladding, down to the $\SiN$ core. The pocket then receives a GaAs device with alignment features complementary to those of the pocket, as seen in Fig.~\ref{fig:Device}a. The GaAs device is then covered with a 1~$\mum$ thick SiO$_2$ cladding layer, as shown in Figs.~\ref{fig:Device}b and c. As shown in Section~\ref{SI_Section_adiabatic} of the SI, an optimized-taper adiabatic mode transformer can be designed to allow highly efficient ($>93~\%$), broadband coupling between the fundamental TE GaAs ridge mode and the $\SiN$ waveguide mode, comparable to values achieved in non-ULLW platforms~\cite{kim_hybrid_2020,elshaari_hybrid_2020}. It is worth noting that our platform may also allow efficient evanescent coupling directly to GaAs nanocavities as demonstrated in Ref.~\onlinecite{katsumi_transfer-printed_2018}. The nanowaveguide geometry features broadband operation, which allows modest Purcell enhancements $(F_p<5)$ that are advantageous for QD spectroscopy and applications in which observation of more than one transition is desired~\cite{Liu2019}. We note also that $F_p\approx5$ may be sufficient to significantly improve the efficiency and indistinguishability of single-photon and pair sources~\cite{Liu2019}.

\begin{figure}[t!]
    \centering
    \includegraphics[width=\columnwidth]{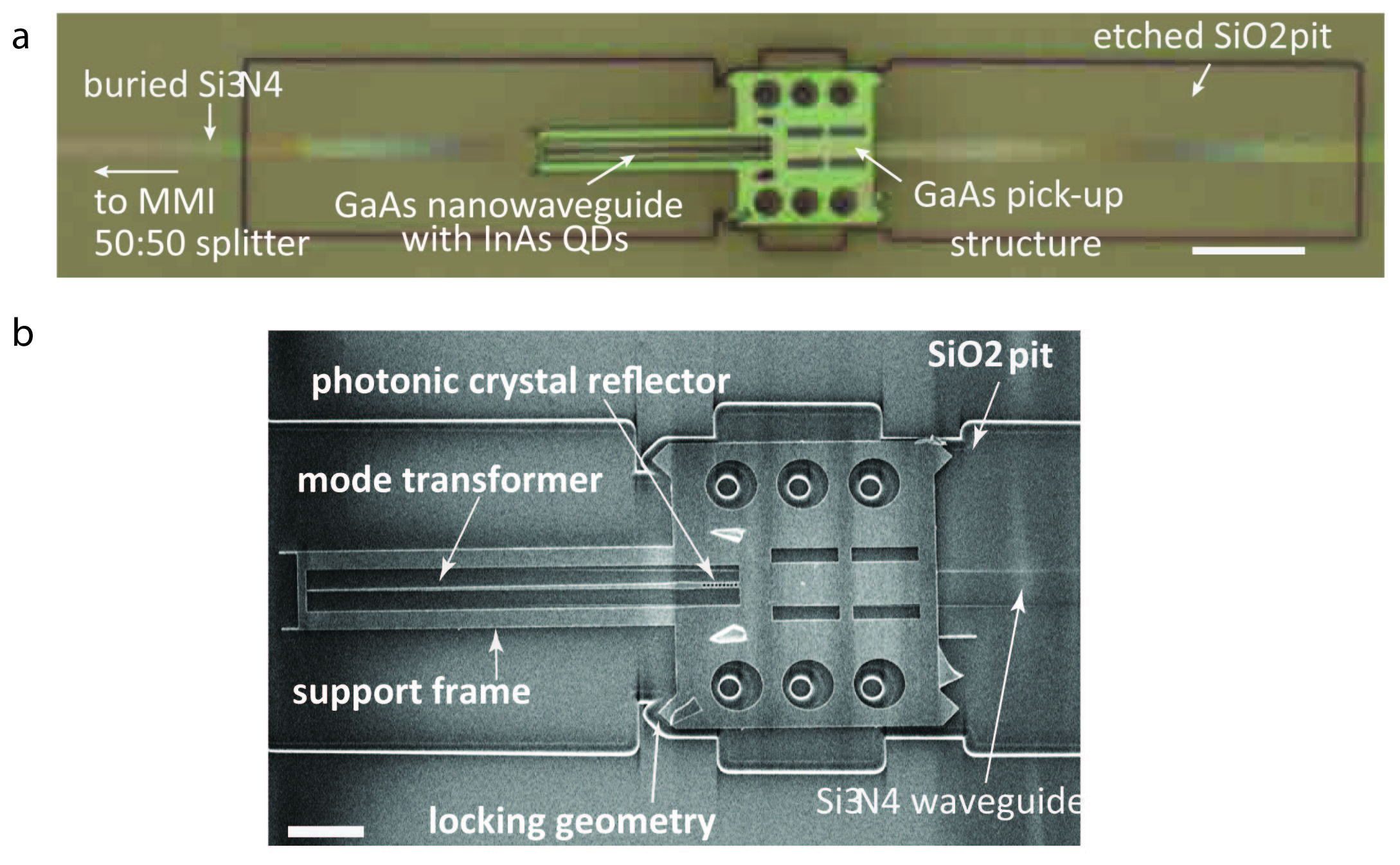}
    \caption{\textbf{a} Optical micrograph of a GaAs / InAs quantum dot single-photon source assembled on a $\SiN$ ultra-low loss waveguide, leading to a 50:50 multimode interference coupler (MMI) power splitter (not shown). The image was taken prior to the top SiO$_2$ cladding deposition. Scale bar: 10~$\mum$ \textbf{b} Scanning electron micrograph of the device prior to deposition of the SiO$_2$ top cladding. Scale bar: 4~$\mum$.}
    \label{fig:Device_fab}
\end{figure}

The hybrid device fabrication is described in \textbf{Methods}. Figure~\ref{fig:Device_fab}a shows an optical microscope image of an assembled GaAs nanowaveguide placed above a buried $\SiN$ ULLW leading into a multi mode interference (MMI) 50:50 splitter. The outline of the etched SiO$_2$ cladding corresponding to the GaAs device placement pit is indicated in the figure. The nanowaveguide geometry, which hosts the quantum dot single-photon emitter, is surrounded by a frame created for mechanical alignment and structural support, and is connected to a pick-up pad that is used for transferring it onto the $\SiN$ chip. A scanning electron micrograph (SEM) in Fig.~\ref{fig:Device_fab}b indicates that there was misalignment between the GaAs device and the $\SiN$ waveguide of $<340$~nm, as well as a tilt angle of $<0.9^\circ$.
 
 \section{Ultra-low loss waveguide characterization}
\label{Section:ULLWG}
To estimate the propagation losses, guides with nominal lengths of 1~m, 2~m and 3~m, implemented as Archimedean spirals~\cite{bauters}, were fabricated and characterized by a single-photon optical time-domain reflectometry (SP-OTDR) technique~\cite{eraerds_photon_2010}. In this technique, short laser pulses in a periodic stream are launched into the ULLW, and photons originating from optical back-scatter along the waveguide are collected and routed towards a single-photon detector. A time-correlator is then used to create a time-trace of back-scattered photon arrival times with respect to a reference clock, and the arrival time can be converted into a distance along the guide. The evolution of the back-scattered light intensity with arrival time provides a direct measure of the signal attenuation along the guide. The experimental setup and details about the measurements and time-to-length conversion are provided in Sections~\ref{SI_Section_OTDR} and ~\ref{SI_Section_ng} of the SI.
\begin{figure}[h]
    \centering
    \includegraphics[width=\columnwidth]{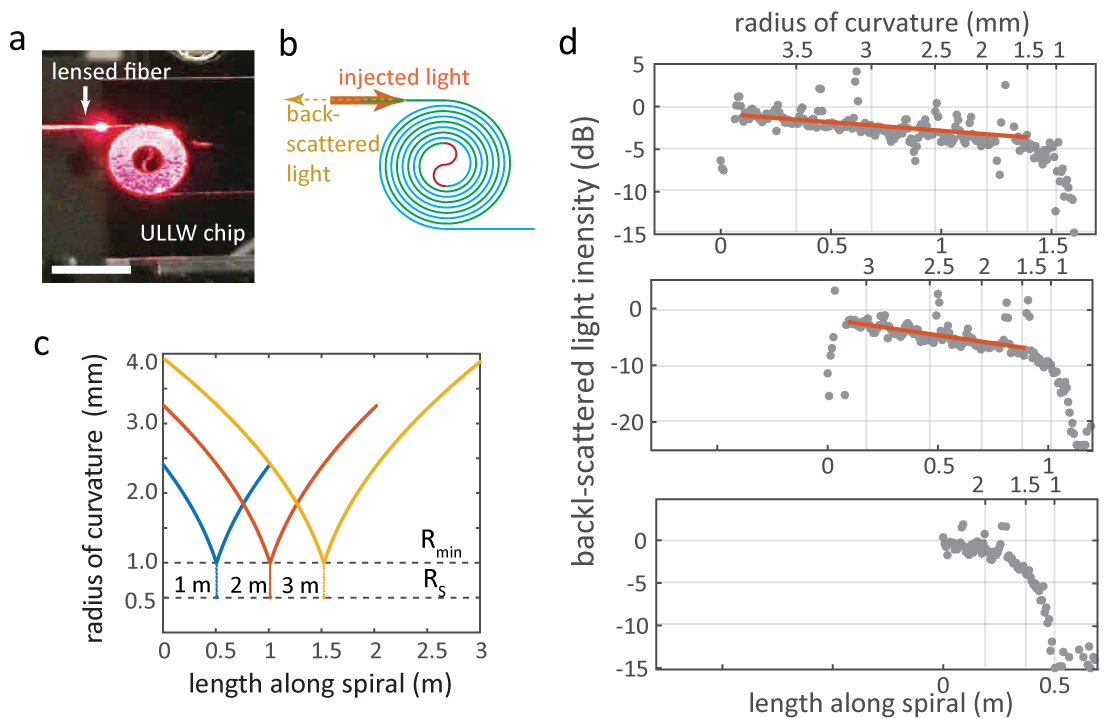}
   \caption{\textbf{a} Photograph of an ultra-low loss waveguide (ULLW) spiral with 1~m length under test. Scale bar: 5~mm. \textbf{b} Schematic of the Archimedean spirals used for loss measurement, composed of inward (green) and outward (blue) spirals, connected by an S-bend (red). In the measurement, laser light is injected into the spiral, and guided back-scattered photons originating along the spiral, collected from the spiral input, are detected in time-domain with a resolution of $\approx200$~ps. \textbf{c} Radius of curvature (RoC) as a function of length for the measured 1~m, 2~m, and 3~m spirals. $R_\text{min}$ and $R_\text{S}$ respectively mark the minimum spiral and S-bend radii. We note the large RoC discontinuity at the S-bend. \textbf{d} Back-scattered light intensity as a function of propagation length and RoC along the 1~m, 2~m, and 3~m spirals, relative to the intensity at the start of each spiral (dots: data; red lines: fits). In each panel, the top and bottom horizontal axes are, respectively, the RoC and length along the corresponding spiral. The spiral length uncertainty is of $<1$~mm, as described in the SI Section~\ref{SI_Section_ng}.}
    \label{fig:ULLWG}
\end{figure}
As shown in Fig.~\ref{fig:ULLWG}c, the Archimedean spirals were designed with a radius of curvature (RoC) that varied continuously going inwards, from a maximum value $R_\text{max}$ - which depended on the total length - to a minimum $R_\text{min}=1000~\mum$ near the center. The inward spiral was followed by an S-bend with $R_{S}=500~\mum$, which transitioned to the outward spiral to the waveguide output. Time-domain reflectivity traces for the three spirals are shown in Fig.~\ref{fig:ULLWG}d, as a function of spiral length  and RoC. All reflectivity curves are approximately linear (in log scale) up to about half of the total spiral lengths. Approximately at the S-bends, the signals drop precipitously. Transmission spectra (not shown) of waveguide-coupled microring resonators with radius $R=500~\mum$ on the same chip did not reveal any resonances, indicating that the signal drop is due to large bend losses at the S-bends. It is also likely that the sharp RoC transition between the spiral and S-bend cause further signal loss. 
To estimate propagation losses in straight ULLWs (bent WGs are not subsequently used in QD integration), linear fits to the OTDR traces were used~\cite{bauters}. The fits were performed for $z$ values from the beginning of the inward spiral to 1~cm before the start of the S-bend, to avoid the abrupt RoC discontinuity. Linear losses for 3~m and 2~m spirals were found to be, respectively, $(1.0 \pm 0.4)$~dB/m and $(2.8 \pm 0.6)$~dB/m. Fits to the 1~m spiral trace did not yield reliable parameters, primarily due to the short extent of the available data.

\section{Triggered single-photon emission}
\label{Section:Triggered}
\begin{figure}[h]
    \includegraphics[width=\columnwidth]{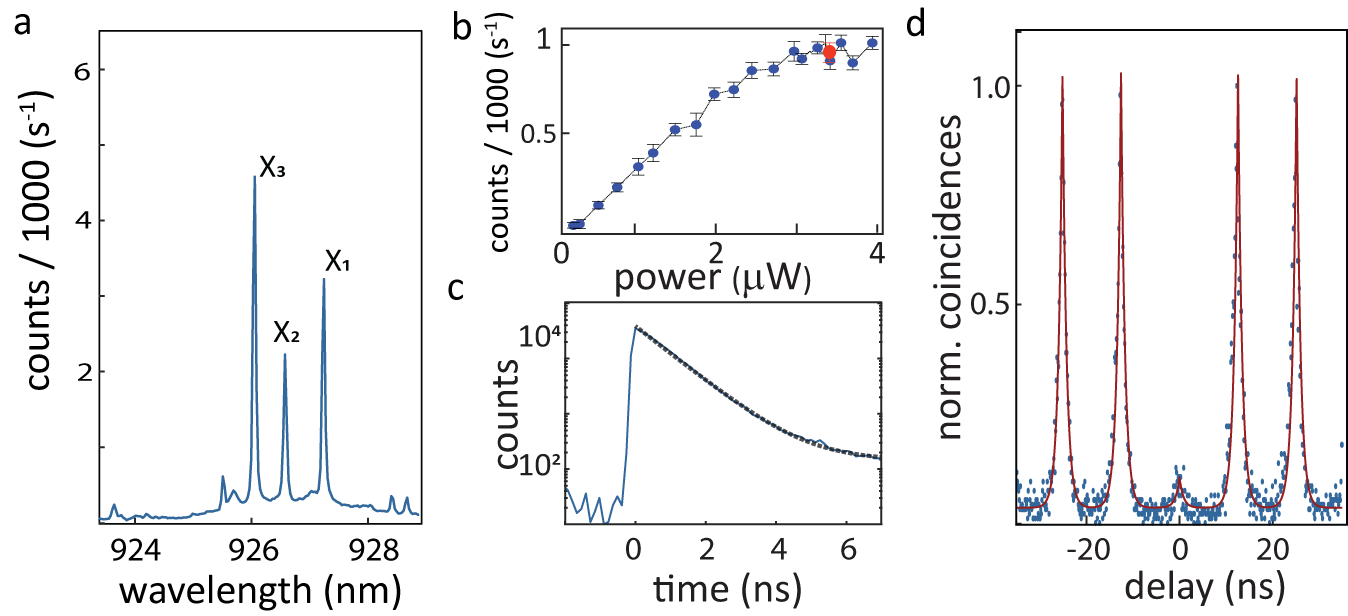}
    \caption{\textbf{a} Quantum dot photoluminescence (PL) spectrum from a hybrid device pumped non-resonantly at $841.5$~nm, showing three transitions from the same QD. \textbf{b} PL intensity for X$_1$ as a function of input power. Red dot: pump level for measurements in \textbf{c} and \textbf{d}. The uncertainties represent $95~\%$ confidence intervals computed from a Lorentzian fit to the QD emission line intensity. The continuous line is a guide to the eye. \textbf{c} Radiative decay trace of the X$_1$ transition measured at saturation, fitted with a biexponential decay function. \textbf{d} Second-order correlation for the $X_1$ line pumped at saturation, showing triggered single photon emission with fitted $g^{(2)}(0)=(0.07 \pm 0.03)$~ns. All uncertainties reported are $95~\%$ fit confidence intervals, corresponding to two standard deviations.}
    \label{fig:Triggered_SPE}
\end{figure}

We next demonstrate triggered single-photon emission from a single QD into a ULLW and characterize its spectral properties and photon statistics at temperatures $< 10$~K. Figure~\ref{fig:Triggered_SPE}a shows the micro-photoluminescence (${\mu}$PL) spectrum obtained for the device in Fig.~\ref{fig:Device}a, pumped from free space with a continuous-wave laser at $841.5$~nm, and collected from the ULLW (details in \textbf{Methods}). The emission lines at $927.21$~nm , $926.57$~nm and $926.02$~nm (labeled as X$_1$, X$_2$ and X$_3$, respectively) were found to be from a single QD via photon-counting cross-correlation measurement. We measured the lifetime of the X$_1$ line by pumping the QD with an 80~MHz, $< 100$~fs  pulse train at $887$~nm. Figure~\ref{fig:Triggered_SPE}c shows the radiative decay time-trace taken at saturation (red dot in Fig.~\ref{fig:Triggered_SPE}b), fit to a double exponential with decay constants $\tau_1= (0.86 \pm 0.01)$~ns and $\tau_2= (2.1 \pm 0.01)$~ns. The slower exponential suggests recombination processes that may adversely affect the coherence of the single photons. To determine the purity of single photon emission, the second-order intensity correlation $g^{(2)}(\tau)$  line was measured in a Hanbury-Brown and Twiss setup. Figure ~\ref{fig:Triggered_SPE}e shows the normalized photon detection coincidences, where a fitted $g^{(2)}(0)=0.07 \pm 0.02$ and decay parameter of $(0.85 \pm 0.02)$~ns was obtained, close to the radiative rate. This shows triggered high-purity, single-photon emission from the QD collected in the ULLW. 

The single-photon count rates produced by the QD pumped into saturation were compared to the 80~MHz pulsed laser repetition rate to yield a measure of the QD-to-ULLW coupling efficiency $\eta_\text{QD-ULLW}$. Assuming $100~\%$ quantum efficiency for the X$_1$ line and discounting all photon losses along the optical path from the $\SiN$ ULLW to the employed superconducting nanowire single-photon detector (SNSPD), we estimate $4~\%\leq\eta_\text{QD-ULLW}\leq7~\%$ (details in the SI). Finite difference time-domain (FDTD) simulations of electric dipoles emitting in a hybrid geometry that approximated the fabricated and tested one indicate that $\eta<31~\%$ could in principle be achieved. As detailed in the SI, the discrepancy between experimental and simulated efficiencies is likely due to sub-optimal QD position and dipole moment orientation inside the GaAs nanowaveguide, though contributions from the misalignment between the latter and the underlying ULLW (evident in Fig.~\ref{fig:Device}b) and other geometrical imperfections were potentially significant. 

 \vspace{0.5cm}

\section{Resonance Fluorescence}
\label{Section:RF}
An additional necessary characteristic for on-chip single-photon sources is high single-photon indistinguishably, which requires the benchmark $T_2=2T_1$ for the quantum emitter coherence time $T_2$, where $T_1$ is the radiative lifetime. Non-resonant excitation of the QD results in an excess of electrons and holes in the host semiconductor and leads to a fluctuating charge environment that inevitably leads to single photons with $T_2<<T_1$. Resonant QD excitation, on the other hand, has been shown to minimize decoherence, allowing the radiative limit to be approached, by avoiding excess environmental charge fluctuations~\cite{Kuhlmann2013a}. An inherent challenge of such a scheme, however, is to sufficiently suppress a pump beam that is resonant with the quantum emitter fluorescence. In free-space-coupled systems, suppression is typically achieved through polarization filtering of the pump before detection~\cite{wang_towards_2019}, though excitation with an orthogonally directed free-space beam~\cite{ates_post-selected_2009} or waveguide~\cite{huber_filter-free_2020} has also been used, and a bi-chromatic pumping scheme has also been recently explored~\cite{he_coherently_2019}. In PICs featuring direct quantum dot resonant illumination with a free-space beam, off-chip polarization filtering before detection has been employed~\cite{dusanowski_near-unity_2019,makhonin}, as well as temporal detection gating~\cite{reithmaier_on_chip_NL_2015}.
In our device and experimental configuration, we observed the resonance fluorescence spectrum collected directly into the ULLW, without polarization filtering or temporal gating. We measured an extinction ratio of $>25$~dB using resonant laser excitation by controlling the polarization of the incident laser alone. This was made possible due to high spatial mode filtering provided by the high aspect ratio  ULLW, which only supports a TE mode, so that the polarization orthogonal to the one supported by the waveguide is highly suppressed. We note that resonance fluorescence has also been observed without polarization filtering in AlN circuits with integrated Ge-vacancy quantum emitters in diamond~\cite{wan_large-scale_2020}.

\begin{figure}[h!]
    \includegraphics[width=\columnwidth]{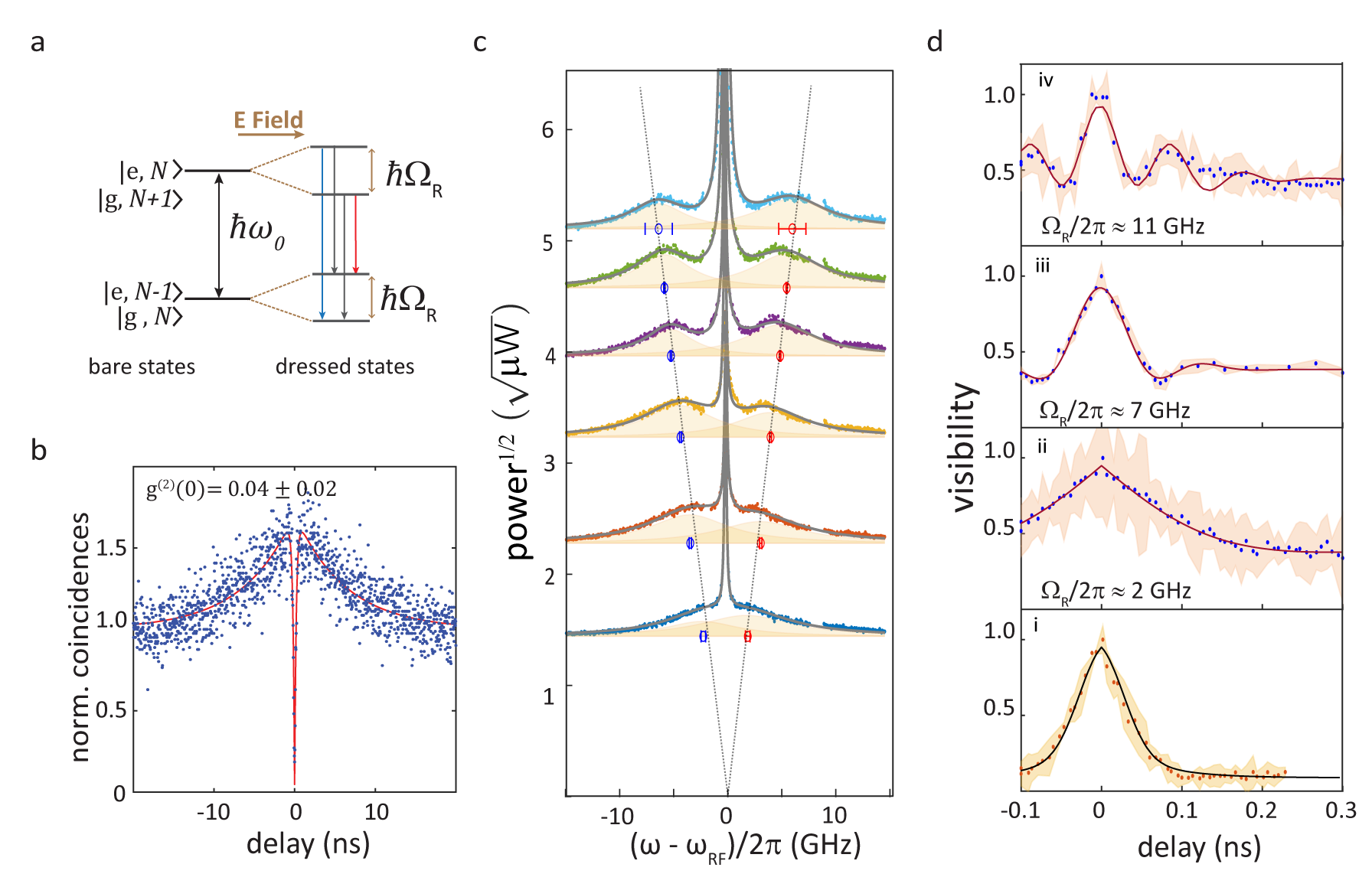}
    \caption{\textbf{a} Energy levels of a two-level system (TLS) driven by a resonant coherent optical driving field. Electric dipole interaction with the driving field splits the TLS `bare' states into two `dressed' states separated by $\hbar\Omega_\text{R}$, producing three emission peaks, represented by different colors. The labels e, g and $N$ respectively correspond to the TLS ground and excited states, and the coherent field average photon number. \textbf{b} Measured second-order correlation and fit (red line) for the QD emission under resonant excitation with $\Omega_{R}/2\pi\approx3.3$~GHz. $g^{(2)}(0)=0.04~\pm~0.02$ indicates single photon emission. \textbf{c} Strong-drive resonance fluorescence spectra for increasing excitation power (colored dots), and corresponding fits (gray continuous lines). The red and blue Mollow side-peak positions obtained from the fits are marked in red and blue symbols respectively. The side-peak energy split increases linearly with the square-root of the excitation power, as indicated by the dotted gray line, from a linear fit to the data. \textbf{d} Interferometric fringe visibility as a function of time delay for QD emission. Panels (ii) to (iv) are for resonance fluorescence, whereas (i) is obtained with quasi resonant (p-shell) pumping at $877.5$~nm. All reported uncertainties correspond to 95~\% fit confidence intervals, corresponding to two standard deviations.}
    \label{fig:RF}
\end{figure}

The resonance fluorescence spectrum of a two-level system varies significantly with excitation intensity. At excitation powers significantly below the saturation level, elastic resonant Rayleigh scattering dominates, and a spectrally narrow emission line is observed~\cite{Matthiesen2012}. At high excitation power, the spectrum features a central resonant peak and two symmetric side-resonances, forming the so-called Mollow triplet~\cite{flagg_resonantly_2009,konthasinghe_coherent_2012}. Waveguide-coupled resonance fluorescence from single quantum emitters has previously been demonstrated in various single-material~\cite{dusanowski_near-unity_2019, makhonin,Thyrrestrup2018,uppu_-chip_2020} and hybrid~\cite{errando-herranz_resonance_2021,wan_large-scale_2020} PIC platforms. In contrast with all this prior work, below we report observation of the Mollow triplet in waveguide-coupled emission, from the same device as measured in the previous Section. The origin of the triplet can be understood from the schematic in Fig.~\ref{fig:RF}a. Two bare states of the quantum dot-field system are split by electric dipole interaction with a strong excitation field, forming a quartet of dressed states. The doubly-degenerate transitions at the resonant energy and the blue- and  red- shifted transitions compose the Mollow triplet. The side-peak splitting is given by the Rabi frequency, $\Omega_\text{R}$, which is proportional to the electric field amplitude.

\vspace{0.5cm}

 To observe resonance fluorescence from our device, a free-space laser beam tuned to the X$_1$ transition in Fig.~\ref{fig:Triggered_SPE}b was used. As detailed in the \textbf{Methods}, control of the pump beam polarization was used to minimize scatter into the ULLW, and a weak non-resonant co-pump was used to gate the resonant emission~\cite{nguyen_optically_2012}. Resonantly driven single-photon emission was first verified via a second-order photon correlation measurement of the resonance fluorescence spectrum. The data, shown in Fig~\ref{fig:RF}b, displays a clear anti-bunching dip, with a fitted $g^{(2)}(0)=0.04~\pm~0.02$, indicating nearly pure single-photon emission. A bunching peak at $\approx3$~ns, however, indicates flickering due to dark state shelving~\cite{Davanco2014} or spectral diffusion~\cite{makhonin}, with a time-scale of $\approx6.4$~ns. Such behavior is likely due to a fluctuating charge environment surrounding the QD, which is ameliorated, though not completely suppressed, by the non-resonant co-pump~\cite{gazzano_effects_2018}. We note also that the X$_1$ transition radiative lifetime $T_1$ was measured to be $T_1 = (0.63 \pm 0.01)$~ns, slightly shorter than previously measured under non-resonant excitation. Such discrepancy is likely due to a slower QD excitation dynamics in the latter case, leading to broadened lifetime traces~\cite{Liu2018}.
 
 Figure~\ref{fig:RF}c shows high resolution resonance fluorescence emission spectra, obtained with a scanning Fabry-Perot interferometer (SFPI), for varying pump powers (details in \textbf{Methods}). The spectra display a sharp Lorentzian central peak and two side-peaks, spaced from the latter by an energy that varies linearly with the excitation field amplitude (square-root of the power), a signature of the Mollow triplet. The sharp central peak includes the elastic contribution of the Mollow spectrum, and scattered resonant pump light. The side-peaks show a slight asymmetry in amplitude and width, which  suggests some detuning between the laser and the transition~\cite{Ulhaq2013}, and spectral diffusion, at time-scales $>T_1$~\cite{konthasinghe_coherent_2012}. Indeed, as shown in Fig.~\ref{SI_Fig:RF_FWHM} of the SI, a model that takes into account QD spectral diffusion~\cite{konthasinghe_coherent_2012} is able to fit the data, yielding $T_2<100$~ps. To confirm and better estimate $T_2$, we use Fourier Transform spectroscopy~\cite{Berthelot2006_NatPhys_motionalnarrowing}. Here, the resonant QD emission was fed into a variable-delay Mach-Zehnder interferometer, and output interference fringe amplitudes were recorded as a function of time-delay. The resulting traces, shown in Fig.~\ref{fig:RF}d, are proportional to the first-order correlation function of the QD light ~\cite{Berthelot2006_NatPhys_motionalnarrowing}, and were fitted to a model~\cite{proux_measuring_2015} that yielded the coherence time $T_2$, as well as the Rabi frequency $\Omega_\text{R}$ (see~\textbf{Methods} and the SI Section~\ref{SI_Section_RF_fits} for details). A reference visibility trace, obtained for non-resonant pumping, is shown in panel \textbf{i} of Fig.~\ref{fig:RF}d. The trace is fitted with a Gaussian, which indicates spectral diffusion, and yields $T_2=(0.053 \pm 0.003)$~ns. Panels \textbf{ii} to \textbf{iv} in Fig.~\ref{fig:RF}d are visibility traces for resonance fluorescence for varying excitation powers, as indicated by the Rabi frequencies. It is worth noting that at the higher powers Rabi oscillations are visible, which are reasonably well reproduced by the model~\cite{proux_measuring_2015}. The corresponding coherence times are \textbf{i} $T_2=(0.10 \pm 0.1)$~ns, \textbf{ii} $T_2=(0.07 \pm 0.01)$~ns and \textbf{iii} $T_2=(0.09 \pm 0.01)$~ns, longer than for the non-resonant excitation values. The coherence dynamics at high powers are better fit with a Gaussian decay, while at lower powers coherence decays exponentially, which indicates prevalence of spectral diffusion~\cite{Berthelot2006_NatPhys_motionalnarrowing}. 

\section{Discussion}
\label{Section:Discussion}
Our work demonstrates for the first time the integration of a quantum emitter single-photon source onto photonic integrated circuits with waveguide losses of $\approx1$~dB/m. These losses, measured at a wavelength of $\approx920$~nm, are more than two orders of magnitude lower than reported in any prior PIC with on-chip quantum emitter single-photon sources, and are substantially lower, by $>3$~dB/m, than the lowest reported for any foundry-compatible quantum PIC to date. We next outline and discuss improvements to achieve the full potential of our integration platform.

Regarding the relatively low  single-photon coupling efficiency into the ULLWs demonstrated here, the main contributing factors include a sub-optimal nanophotonic design and quantum dot positioning and, principally, dipole moment orientation within the GaAs device. While various techniques have been developed to solve the latter issues~\cite{Liu2018,schnauber2019indistinguishable} the implemented photonic design featured two factors that fundamentally lead to lower efficiencies. First, the choice of a waveguide geometry imposes a limit on the QD coupling to guided, as opposed to radiative, waves~\cite{Davanco2017}. Second, the linear taper profile of the mode transformer leading to the $\SiN$ waveguide displayed sub-optimal efficiency, though, as discussed in Section~\ref{fig:Device} and the SI, superior designs with significantly higher efficiencies can  be implemented. While a waveguide geometry may be desirable for broadband operation, evanescently coupled microcavities are a viable alternative towards achieving higher overall coupling efficiencies~\cite{katsumi_transfer-printed_2018} and are the subject of future work. An additional advantage of this approach is that a high Purcell radiative rate enhancement, achieved through coupling to the resonant mode, can bring the quantum emitter's lifetime $T_1$ closer to the radiative limit $T_2=2T_1$, given a coherence time $T_2$ that is sufficiently unaffected by nanofabrication, thereby improving indistinguishability~\cite{Liu2018,Liu2018b}. On the other hand, a single quantum dot exhibits various excitonic transitions over a relatively wide spectral range, which may be used for desirable functionalities beyond triggered single-photon emission. For instance, polarization-entangled photon pairs may be generated from the biexciton-exciton cascade~\cite{Liu2019}, where the two states are typically split by $\approx1$~nm. These entangled photon states, when captured into an integrated photonic circuit, could present interesting opportunities for quantum information processing on a chip. 

Regarding collection of resonance fluorescence with higher pump suppression, fine control of the QD orientation will likely be necessary. Control of the resonant pump polarization was shown here to effectively minimize scatter into the ULLW. Keeping in mind that only the QD dipole moment component that is transverse to the ULLW couples to it, the QD must be oriented such that the (optimally polarized) pump maximizes resonant QD emission into the ULLW. The QD must have a sufficiently large dipole moment component along the pump polarization to excite QD emission above the scattered light level. In principle, though, with proper design of components, a higher degree of pump suppression can be achieved. While it is unclear what factors contribute most to scatter from the free-space pump into the ULLW, it is likely that fabrication imperfections are to blame, which brings an undesirable degree of uncertainty to the problem. As an alternative, waveguide-based resonant pumping may provide more controllable means of minimizing waveguided pump scatter~\cite{uppu_-chip_2020}.

The broad linewidths observed even upon resonant excitation, due to large spectral diffusion and dephasing, limited our ability to coherently control the quantum dot and demonstrate indistinguishable single-photons. In particular, the need to co-pump the quantum dot non-resonantly with above-band light most likely contributed to an increase of the inhomogeneous linewidth particularly at higher resonant excitation~\cite{gazzano_effects_2018}. It is unclear whether any of the fabrication steps were ultimately responsible for the large spectral diffusion in our devices, since the quantum dots were not characterized pre-fabrication. Screening the QD population prior to fabrication may allow identification of QDs with narrower linewidths. Deterministic positioning of single QDs within nanofabricated geometries, at sufficient distances from etched sidewalls, has been shown to be at least beneficial in preserving emission properties~\cite{Liu2018b,schnauber2019indistinguishable}.

Regarding our passive photonic circuits, lower propagation losses may be achieved by employing blanket nitride growth, etch, and annealing techniques~\cite{Pucket2020}, as well as transverse magnetic (TM) field designs \cite{liu_720_2021}. At the same time, we anticipate that a variety of on-chip passive components already demostrated in this platform, including spiral delay lines~\cite{Huffman2017}, filters~\cite{Huffman2018}, and couplers and switches~\cite{moreira_compact_2015},  can be further optimized for lower insertion losses. 

Implementing all of the measures above - improving the QD-to-waveguide coupling efficiency and enhancing single-photon indistinguishability via nanophotonic design and deterministic QD positioning, and further minimizing propagation and insertion losses in passive on-chip components - will bring us closer to fully chip-integrated systems implementing practical Boson sampling and related photonic quantum information tasks with quantum advantage. We note further that the ultra-low propagation losses demonstrated here may already allow the implementation of on-chip delays for time-demultiplexing of a single quantum emitter single-photon source, to produce spatially multiplexed photons for Boson sampling similar to that demonstrated with free-space optical delays in ref.~\onlinecite{Wang2019}.

In conclusion, our results indicate high prospects for the utilization of quantum emitters as on-demand sources of single-photon in ultra-low loss, $\leq1$~dB/m , photonic integrated circuits, which may prove essential for the creation of scaled photonic quantum information systems on-chip.



\clearpage
\noindent \large{\textbf{Methods}}\\
\medskip

\noindent \small{\textbf{Uncertainty reporting}} 
Wherever unspecified in the text, reported uncertainties are 95~\% confidence intervals, corresponding to two standard deviations, resulting primarily from Type A evaluations of least-squares fits of models to data. We report other details of uncertainty evaluation as relevant.

\noindent\small{\textbf{Estimation of misalignment between GaAs and $\SiN$ waveguides}}
To estimate the misalignment between the $\SiN$ and GaAs waveguides in the SEM of Fig.~\ref{fig:Device}b, we calibrate the image pixel size using reference positions produced by electron-beam-lithography on the GaAs device. We then measure pixel distances between $\SiN$ waveguide and GaAs support frame at various locations to determine physical distances and tilt angles. Although the uncertainty is expected to be negligible, because we do not evaluate the uncertainties related to edge thresholds, we provide conservative estimates of $<340$~nm and $<0.9^\circ$ for the lateral displacement and tilt angle, respectively.

\noindent \small{\textbf{Device Fabrication}} 
Device integration involves fabricating III-V semiconductor single photon emitters in a tab-released membrane structure and employing a pick-and-place technique~\cite{mouradian_scalable_2015,wan_large-scale_2020} to place the emitter in pockets etched in the $\SiN$ waveguide upper oxide cladding. Alignment is achieved in the x-y plane using etched mechanical features in the semiconductor and waveguide upper cladding oxide pocket. Fabrication of the $\SiN$ chip and the GaAs/QD devices was done in two separate runs. For the passive, ULL circuit, low pressure chemical vapor deposition (LPCVD) $\SiN$ was deposited on a 100~mm silicon wafer with a 15~$\mum$, thermally grown SiO$_\text{2}$ layer. Waveguides were patterned with a deep-ultraviolet (DUV) stepper and dry etched using an inductively coupled plasma (ICP) reactive-ion etcher (RIE) with CHF$_\text{3}$/CF$_\text{4}$/O$_\text{2}$ chemistry. A $\approx 1~\mu\text{m}$ layer of SiO$_\text{2}$ was deposited by plasma enhanced chemical vapor deposition (PECVD) using liquid tetraethoxysilane (TEOS) as a precursor of Si, followed by a high temperature anneal and chemical mechanical polishing (CMP) for planarization.  Optical lithography was then used to define placement pits for the GaAs devices, aligned to buried $\SiN$ waveguides. The placement pits were etched $\approx500$~nm deep into the top SiO$_\text{2}$ cladding. To better accommodate the QD devices, the pits were further trimmed with an additional optical lithography step followed by a buffered oxide etch (BOE). The visible fringes along the buried waveguide in Fig.~\ref{fig:Device_fab}a show evidence of non-uniform SiO$_2$ removal from above the $\SiN$, and, potentially, also etching of the $\SiN$.
GaAs devices were fabricated from an epitaxially grown stack consisting of a 190~nm thick GaAs layer containing InAs QDs at the center, on top of a 1~$\mu\text{m}$ Al$_{0.7}$Ga$_{0.3}$As sacrificial layer. Prior to fabrication wide-field illumination photoluminescence imaging confirmed the presence of high density quantum dots emitting in the 900 nm band, with individual quantum dots addressable through a combination of spatial and spectral filtering during subsequent device characterization. Electron-beam lithography followed by Cl$_2$/Ar ICP etching was used to define the devices on the epi-wafer, and hydrofluoric acid was used to remove the sacrificial layer. This process resulted in free-standing GaAs devices that could be picked up with a tungsten probe and placed onto the etched pits on the ULLW chip~\cite{wan_large-scale_2020}. The GaAs devices and placement pits had triangular locking geometries (indicated in Fig.~\ref{fig:Device} a) that enable sub-micron alignment to be achieved. The successful integration of the GaAs devices was confirmed using optical microscope as well as scanning electron microscope prior to deposition of the top SiO$_2$ cladding (see SI for details on estimating the device alignment). After device placement into the etched pits, PECVD was used to deposit a 1$\mum$ SiO$_2$ film over the entire chip. This step created a SiO$_2$ upper cladding for the GaAs devices. Before testing, diced chip facets were polished such that the waveguide ends of the spirals were accessible via end-fire coupling. 

\medskip
\noindent \textbf{Cryogenic Photoluminescence measurements} \\
 The fabricated devices were measured in a closed-cycle Helium cryostat at temperatures $< 10$~K. The sample was imaged from the top, with a micro-photoluminescence (${\mu}$PL) setup implemented just above an optical window at the cryostat chamber top~\cite{Davanco2017,schnauber2019indistinguishable}. Optical excitation of the QDs in the GaAs devices was also done from the top, with laser light focused to a spot of $\approx1~\mum$ diameter. Quantum dot emission coupled to the ULLWs was collected using a lensed optical fiber mounted on a nanopositioning stage that could be aligned to WG facets at the polished edge of the hybrid chip. The results shown here were obtained from devices that included 50:50 MMI splitters, as shown in Fig.~\ref{fig:Device_fab}. Figure~\ref{SI_Fig:MMI_PL} in the SI shows ${\mu}$PL spectra produced by one of the fabricated devices under $845$~nm continuous wave (CW) laser pumping, collected separately from the two MMI output ports.

\noindent \textbf{Triggered single-photon emission measurements}
We measured the lifetime of the X$_1$ line upon excitation with a $< 100$~fs, 80~MHz pulsed laser at $887$~nm. The emission was filtered using a $\approx500$~pm bandwidth fiber coupled grating filter having efficiency of $\approx50~\%$ and the photon counts were detected with a superconducting nanowire single photon detector (SNSPD). 

To determine the purity of single photon emission, the intensity autocorrelation for the exciton line was measured using two SNSPDs in a Hanbury-Brown and Twiss configuration. Figure~\ref{fig:Triggered_SPE}e shows the normalized photon detection coincidences, measured with a 128~ps bin size, for the $X_1$ line pumped at saturation (red dot in Fig.~\ref{fig:Triggered_SPE}b, top). The data was fitted with a two-sided exponential decay and a $g^{(2)} (0)$ value of $0.07 \pm 0.02$ and decay parameter of $(0.85 \pm 0.02)$~ns was obtained, close to the radiative rate. This shows triggered high-purity single photon emission from the QD collected in the ULLW. 

\noindent \textbf{Resonance fluorescence measurement}
To observe resonance fluorescence from our device, free-space excitation was used once again, with a laser beam tuned to the X$_1$ transition in Fig.~\ref{fig:Triggered_SPE}b. Polarization control of the excitation beam allowed us to suppress scattered pump light into the $\SiN$ waveguide by as much as $\approx25$~dB while monitoring the signal on a grating spectrometer. In order for the resonance fluorescence to be observable however, it was necessary to co-excite the QD with a weak non-resonant laser at $\approx841$~nm~\cite{nguyen_optically_2012}. While the non-resonant laser alone was sufficiently weak to produce negligible photon emission counts for all resonant laser powers, it enhanced the resonance fluorescence light by as much as $\approx10$ times.

The Mollow triplet spectra shown in Fig.~\ref{fig:RF}c were obtained by filtering QD emission collected from the ULLW with a scanning Fabry-Perot interferometer (SFPI)  with free-spectral range of $40$~GHz and finesse of $\approx200$. At different resonant excitation powers, the intensity of the non-resonant co-pump was optimized to increase the resonant emission count. A $\approx200$~GHz bandwidth fiber-coupled grating filter preceding the SFPI eliminated non-resonant laser light while allowing the complete resonance fluorescence spectrum to be measured. The Mollow triplet spectra were fit, through a nonlinear least-squares method, with a function that included three Lorentzians peaks, corresponding to the center and two side-peaks of the incoherent Mollow triplet spectrum- and an additional, sharp central Lorentzian to account for the coherent resonance fluorescence signal and pump scatter~\cite{konthasinghe_coherent_2012}. The spectral locations of the side-peaks (with 95~\% fit confidence intervals) are plotted as a function of pump power in Fig.~\ref{fig:RF}c. 

A physical model of the Mollow triplet that included effects of laser detuning and QD spectral diffusion~\cite{konthasinghe_coherent_2012} was also used to fit the data, yielding the $T_2<$100~ps estimate given in the main text. Plots of the fits and extracted parameters are shown in the SI Section~\ref{SI_Section_RF_fits}.

\noindent \textbf{Fourier-transform spectroscopy}
For Fourier-transform spectroscopy, QD emission resonant with the pump laser was passed through a Mach-Zehnder interferometer (MZI) with variable delay, then detected with an SNSPD. The MZI delay was scanned to yield an interferogram that corresponded to the first-order correlation function of the QD emission, from where the QD coherence time $T_2$ can be extracted~\cite{proux_measuring_2015}. In our experiment, the MZI was tuned to a discrete number of delay values between -0.1 ns and 0.3 ns. At each point, the MZI delay stage was dithered 5 times with an amplitude of $2~\mum$, giving sufficient time for the system to stabilize. Interference fringes from the latest dither were recorded and the visibility $V=\left(I_\text{max}-I_\text{min}\right)/\left(I_\text{max}+I_\text{min}\right)$, where $I_\text{max, min}$ are the maximum and minimum fringe intensities, was calculated at each point.

\noindent \textbf{Acknowledgements}
We thank Edward Flagg from West Virginia University for helpful discussions regarding the resonance fluorescence results. A.C. acknowledges support under the Cooperative Research Agreement between the University of Maryland and NIST-PML, Award no. 70NANB10H193. J.D.S. acknowledges the program of quantum sensor core technology through IITP (MSIT Grant No. 20190004340011001).

\noindent \textbf{Author Contributions}
M.D. performed electromagnetic design of the hybrid devices. J.C., D.E., R.M., D.J.B., K.S. and M.D. conceptualized the hybrid chip pick-and-place assembly method. J.D.S. provided the molecular beam epitaxy-grown quantum dot sample. R.M. designed and fabricated the $\SiN$ photonic circuit chip. H.L. devised and performed post-fabrication adjustments to the $\SiN$ on-chip devices. B.G. fabricated the GaAs devices. J.C. and H.L. performed pick-and-place assembly of the hybrid device. V.A. provided superconducting nanowire single-photon detectors. A.S. performed the waveguide propagation loss and single quantum dot characterization measurements, analyzed the data and produced figures for the manuscript.  A.S. and M.D. wrote the manuscript, with input from all authors. M.D., K.S., D.J.B., D.E and J.C. supervised the project. All the authors contributed and discussed the content of this manuscript.

\noindent \textbf{Additional Information} Correspondence and
requests for materials should be addressed to M.D.

\noindent \textbf{Competing financial interests} The authors declare no competing financial interests.

\setcounter{figure}{0}
\setcounter{equation}{0}
\setcounter{section}{0}

\makeatletter
\renewcommand{\thetable}{S\arabic{table}}
\renewcommand{\thefigure}{S\arabic{figure}}
\renewcommand{\@biblabel}[1]{#1. }
\renewcommand{\@dotsep}{500}
\renewcommand{\@pnumwidth}{0em}
\renewcommand{\l@figure}[2]{
\@dottedtocline{1}{1.5em}{2em}{Figure #1}{}\vspace{15pt}}
\renewcommand*\arraystretch{1.25}

\clearpage
\setcounter{page}{1}

\begin{center} {{\bf \large Supplementary Information to "Triggered single-photon generation and resonance fluorescence in ultra-low loss integrated photonic circuits"}}\end{center}

\section{Low-loss integrated quantum photonic platforms}
\label{SI_Section_ULLW}
Table~\ref{SI_table:ULL_Q_platforms} shows the lowest reported propagation losses, to the best our knowledge, for various current integrated quantum photonic platforms. Also indicated are whether the demonstrated devices included on-chip sources, and of what type. The current work not only exhibits the lowest losses for foundry-compatible integrated quantum photonic platform, but also combines it with the triggered single-photon emission made possible by integration with a single quantum emitter. 

\begin{table}[h!]
\centering
\resizebox{\textwidth}{!}{
\begin{threeparttable}
\caption{Reported propagation losses for various photonic integration platforms used in quantum photonics demonstrations to date. SFWM: spontaneous four-wave mixing; SPDC: spontaneous parametric down-conversion. PPLN: periodically-poled lithium niobate}
\label{SI_table:ULL_Q_platforms}
\begin{tabular}{|c|c|c|c|c|c|c|c|}
\hline
\textbf{Integration platform} & \textbf{Losses (dB/m)}        & \textbf{Wavelength (nm)} & \textbf{Source type} & \textbf{On-/off-chip source}& \textbf{Foundry-compatible$^*$} & \multicolumn{1}{c|}{\textbf{Reference}} \\ \hline
Laser-written SiO$_2$         & $<30$                         & 1550                     & no on-chip source    & Off-chip        &No                   &         \cite{ceccarelli_low_2020}                                 \\ \hline
Laser-written SiO$_2$         & $\approx 50$                  & 800                      & no on-chip source    & Off-chip        &No                   &         \cite{spagnolo_experimental_2014}                                \\ \hline
UV-written silica-on-Si       & $\approx15$                   & 1550                     & SFWM                 & On-chip         &No                   &         \cite{posner_high-birefringence_2018}                               \\ \hline
Double-stripe $\SiN$          & $\approx20$                   & 1550                     & SFWM                 & On-chip         &Yes                &         \cite{taballione_88_2019}                             \\ \hline
Hydex                         & $\approx5$                    & 1550                     & SFWM                 & On-chip         &Yes                &         \cite{moss_new_2013}                             \\ \hline
$\SiN$ ridge                  & $\approx5$ $^\text{a}$        & 1550                     & SFWM                 & On-chip         &Yes                &         \cite{ramelow_silicon-nitride_2015}                           \\ \hline
AlN ridge                     & $\approx100$ $^\text{b}$      & 1550                     & SPDC                 & On-chip         &Yes                &         \cite{guo_parametric_2017}                           \\ \hline
Silicon-on-insulator          & $\approx100$                  & 1550                     & SFWM                 & On-chip         &Yes                &         \cite{ma_silicon_2017}                           \\ \hline
AlGaAs-on-Insulator           & $\approx20$                   & 1550                     & SFWM                 & On-chip         &Yes                &         \cite{steiner_ultrabright_2021}                           \\ \hline
Ti:PPLN                       & $\approx 1.6 / \approx 2.2$   & 890 / 1320               & SPDC                 & On-chip         &No                   &         \cite{luo_direct_2015}                           \\ \hline
Ti:LiNbO$_3$                  & $\approx10$                   & 1550                     & SPDC                 & On-chip         &No                   &         \cite{luo_nonlinear_nodate}                          \\ \hline
GaAs/Si$_3$N$_4$              & $\approx$100                  & 900                      & quantum emitter      & On-chip         &Yes                &         \cite{Davanco2017}                          \\ \hline
GaAs/Si$_3$N$_4$              & $\approx~1$                   & 900                      & quantum emitter      & On-chip         &Yes                &         \textbf{This work}                          \\ \hline
\end{tabular}
\begin{tablenotes}
   \item[a] Estimated from measured microring resonator intrisic quality factors, $Q_i\approx7\times10^7.$ \\
      \item[b] Estimated from measured microring resonator intrisic quality factors $Q_i\approx4\times10^5$.\\
      \item[*] Having at least \emph{passive} photonic circuits that can be produced in a semiconductor foundry.  
\end{tablenotes}
\end{threeparttable}
}
\end{table}

\section{Single-photon Optical Time-domain Reflectometry (SP-OTDR) setup}
\label{SI_Section_OTDR}

\begin{figure}[h]
    \centering
    \includegraphics[width=0.9\columnwidth]{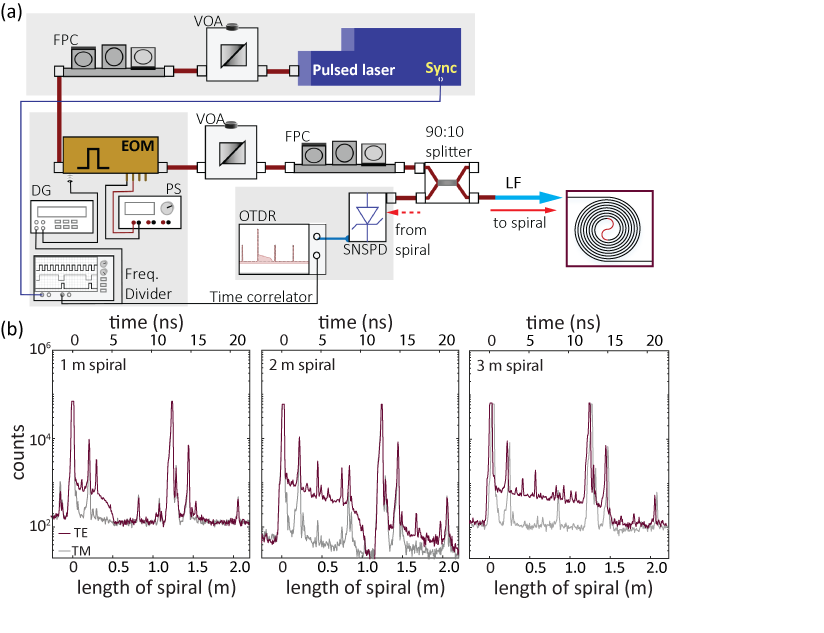}
    \caption{(a) Schematic of experimental setup for single-photon optical time-domain reflectometry (SP-OTDR) measurements. FPC: fiber polarization controller; VOA: variable optical attenuator; EOM: electro-optic modulator; PS: power supply; DG: delay generator. LF: Lensed Fiber. (b) SP-OTDR signal for back-scattered light from 3~m Archimedes spiral, upon injection of $<100$~fs laser pulses. Exponentially decaying counts are observed as a function of time after pulse at zero delay for TE polarized excitation (red), indicating propagation in the spiral. Data for TM polarization is shown in gray, indicating much larger propagation losses, as to be expected given that the waveguide does not support a TM mode.}
    \label{SI_fig:OTDR_setup}
\end{figure}

Figure~\ref{SI_fig:OTDR_setup}(a) shows the schematic of the experimental setup used for our SP-OTDR measurements. An 80~MHz mode-locked fiber laser was used to produce $<100$~fs pulses centered at a wavelength of $\approx$930~nm. An electro-optic modulator (EOM) synchronized to the laser was used to  attenuate its output by $>20$~dB, allowing a single pulse to pass at a period of more than 100~ns. This was done in an attempt to limit the observation of stray reflected pulses within the time-window corresponding to the spiral lengths. The modulated signal was passed through a 90:10 fiber splitter, then coupled to a lensed optical fiber connected to the 10~\% port of the latter. The lensed fiber both launched the pulses into the on-chip ULLWs and collected the reflected light. The latter was then routed to superconducting nanowire single-photon detectors (SNSPDs) via the 90:10 splitter. The time correlator had a specified temporal resolution of 4~ps. Coupling into the TE $\SiN$ waveguide mode was achieved by maximizing the reflected signal by varying the polarization of the injected light with a fiber polarization controller (FPC) placed before the 90:10 splitter. Minimizing the reflected signal (consistent with TM-polarization coupling) allowed us to obtain a mostly flat noise background, with spurious peaks which were also visible on the TE polarization curve. Such peaks originate from polarization-independent interfaces (e.g., fiber connectors and the chip facet) or scatterers along the probing light path, or are pulses that were insufficiently attenuated by the EOM. The TE mode and TM-polarization (background) reflectivity curves for the three  measured spirals are shown in Fig.~3(b) in the main text. Subtracting the background signal from the TE reflectivity curves allowed us to reduce some of these features. The fits reported in Fig.~3(d) were done on such background-subtracted data, also displayed in the figure. To find the start of each spiral in time, we note that each spiral trace starts with a tall peak due to the chip facet. We select as the spiral start time the point at the onset of the facet peak, where the intensity is 10~dB higher than the preceding (background) intensity level.

The optical path traversed by the photons back-scattered at any point is twice the length of the spiral at the point.  Thus the total length traversed by the photons, for calculation of loss, is twice the length of the spiral or the length shown as horizontal axis in Fig~\ref{SI_fig:OTDR_setup}(b). Since the high-aspect ratio ULLW only supports the TE mode, the TM mode doesn't propagate in the waveguide and is reflected from the chip facet. Thus the OTDR for TM polarization also serves as the reference signal. The common peaks among the signal for TE and TM polarization, as well as after other trains of pulses, indicates that these originate from elsewhere. These were identified as reflection from other optical components used in the setup.

\section{Determination of the ULLW group index}
\label{SI_Section_ng}
\begin{figure}[h!]
    \centering
    \includegraphics[width=0.5\columnwidth]{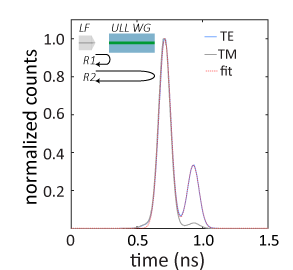}
    \caption{Time-of-flight traces of reflected short pulses from an $\approx$21~mm long straight ultra-low loss waveguide (ULLW), launched from a lensed optical fiber (LF) in endfire probing configuration. The two peaks correspond to first ($R_1$) and second ($R_2$) facet reflections, as indicated in the inset. Curves for TE and TM polarization are shown, evidencing considerably larger propagation losses for the latter, as expected.}
    \label{SI_fig:OTDR_ng}
\end{figure}

The time-axis in our SP-OTDR measurements can be converted into a propagation length along the ULLWs though the relation
\begin{equation}
\label{SI_eq_ng}
    \Delta L = \frac{c_0\Delta t}{2n_g},
\end{equation}
where $c_0$ is the velocity of light in vacuum, $\Delta L$ is the length propagated in a time interval $\Delta t$, and  $n_g$ is the group index of the waveguide. To estimate $n_g$ in the ULLWs, we performed SP-OTDR measurements on a straight waveguide fabricated on the same chip as the spirals studied in the main text. Figure~\ref{SI_fig:OTDR_ng} shows time-dependent reflectivity traces obtained my maximizing and minimizing the insertion loss (IL) via the input polarization. Both traces were obtained with the same integration time of 60~s. Two peaks separated by $\approx200~$ps are apparent in the figure. Because the ULLW was designed to be close to single-mode, we assign the low IL (blue) curve to the fundamental transverse-electric (TE) mode, whereas the gray curve is assigned to the first transverse-electric (TM) mode. Because the first peaks for both the TE and TM polarizations have approximately the same amplitude, we assign such peaks to  reflection at the waveguide facet, at the edge of the chip just after the lensed fiber. The next peaks, spaced by $\approx200$~ps, are assigned to the first reflection at output end of the ULLW. This assignment is based on the following reasoning. 

A two-Gaussian fit to the TE polarization trace (continuous curve in Fig.~\ref{SI_fig:OTDR_ng}) gives a delay $\Delta t=(-206.0\pm0.9)$~ps, amplitude ratio $R_2/R_1=0.260\pm0.004$, and a width ratio $\sigma_2/\sigma_1=1.040\pm0.005$ between the first and second peaks. The uncertainties here are 95~\% confidence intervals from the nonlinear fit, corresponding to two standard deviations. We note that the pulse widths are limited by the temporal resolution of our TCSPC system. 

Assuming that the Gaussian beam produced by the lensed fiber refracts at the interface following simple Fresnel relations for normal incidence, the reflection coefficients at the first and second facets, $r_1$ and $r_2$, are
\begin{equation}
    r_{1,2} = \pm\frac{n_o-n_i}{n_o+n_i},
\end{equation}
where $n_i$ and $n_o$ the refractive index outside and inside the device dielectric. We take $n_o=1.0$ for air, and $n_i\approx n_{\text{eff}}=1.466$, the TE mode effective mode index, computed with the Finite Element Method. The Fresnel transmission coefficients at the two facets are
\begin{equation}
    t_{1,2} = 2\frac{n_{o,i}}{n_o+n_i}.
\end{equation}
The reflectance from the first facet back into the fiber is $R_1=\eta_1|r_1|^2$. We have introduced $\eta_1$ to represent coupling losses between the free-space beam reflected at the facet and the fiber mode. The reflectance from the second facet back into the fiber is
\begin{equation}
    R_2=\eta_2^2\left| r_2 t_1 t_2\right|^2,
\end{equation}
accounting for Fresnel transmission at the first facet, reflection at the second facet and transmission back through the first facet. The factor $\eta_{2}$ is introduced to represent modal coupling losses between the fiber and the TE waveguide mode. With this,
\begin{equation}
\label{SI_eq_R2R1}
    \frac{R_2}{R_1}=\frac{\eta_2^2}{\eta_1}\left|t_1 t_2\right|^2 \approx  0.9 \frac{\eta_2^2}{\eta_1}
\end{equation}
The fiber-to-ULLW coupling efficiency is estimated as $\eta_{2}\approx0.55$, as described in Section~\ref{SI_section:coupling_eff}, from insertion loss measurements on straight waveguides fabricated on the sample chip as the spirals. Replacing in eq.~(\ref{SI_eq_R2R1}), we obtain $R_2/R_1\approx0.27/\eta_1$. Comparing with $R_2/R_1\approx0.3$ obtained from the experimental fits above, we estimate $\eta_1\approx0.9$, which is a reasonable value. 

Next, we used an optical microscope with an encoded sample stage to determine the length $L$ of the measured waveguide. The stage encoder had a specified resolution of better than 100~nm. Using the coordinates of the four corners of the rectangular waveguide geometry at 100$\times$ magnification, we obtain $L\approx20.9$~mm over four measurements. The Student t-distribution 95~\% confidence interval for the four measurements is $\approx0.6~\mum$. Using the $n_g=1.51$ value computed with the finite element method, we calculate an expected delay $\Delta_t\approx210$~ps between the first two reflected pulses, which is reasonably close to the $\approx206$~ps value obtained from the time-trace fit (i.e., it corresponds to a length difference of $\approx1$~mm, which we take to be our length uncertainty in Fig~\ref{fig:ULLWG}d of the main text). Taking the measured waveguide length and fitted delay between the pulses into eq.~(\ref{SI_eq_ng}), we obtain $n_g=1.482\pm0.006$. The uncertainty reported here is obtained by propagating uncertainties for $\Delta t$ and $L$ in eq.~(\ref{SI_eq_ng}). For $\Delta t$, we use the 95~\% fit confidence interval reported above as the uncertainty, whereas for $L$ we use the Student t-distribution 95~\% confidence interval.

\clearpage
\section{Single-photon coupling efficiency}
\label{SI_section:coupling_eff}
We estimate the single-photon coupling efficiency $\eta_\text{QD-ULLWG}$ from the QD into the ULLW in our device by pumping the former into saturation with an 80 MHz pulsed laser at 880~nm, assuming 100~\% quantum efficiency for the emission, and comparing with the single-photon detection rate, after taking into account all the losses along the path to the detectors. Specifically the ULLW-coupled QD light goes through the MMI with efficiency $\eta_\text{MMI}$, and is collected by the lensed fiber with efficiency $\eta_\text{LF}\cdot\eta_\text{facet}$.  In the latter expression, the first factor is just due to the fiber itself, including the FC/APC connector it featured on one end, and the second the fiber-to-waveguide coupling efficiency at the facet. Fiber-coupled photons are subsequently passed though a $\approx0.5~$nm bandpass grating filter, with efficiency $\eta_\text{filter}$, then routed to an SNSPD with detection efficiency $\eta_\text{SNSPD}$ via an optical fiber path with efficiency $\eta_\text{fiber}$.The detection rate at the SNSPDs is given by $R_\text{det.}=\eta_\text{QD-ULLWG}\cdot\eta_\text{MMI}\cdot\eta_\text{facet}\cdot\eta_\text{LF}\cdot\eta_\text{filter}\cdot\eta_\text{fiber}\cdot\eta_\text{SNSPD}$.

\begin{figure}[h]
    \centering
    \includegraphics[width=0.5\columnwidth]{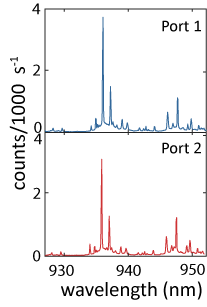}
    \caption{Quantum dot (QD) photoluminescence collected from the two output ports of a ULLW based 50:50 MMI splitter in the hybrid sample. Here, the QD was pumped at 845~nm. }
    \label{SI_Fig:MMI_PL}
\end{figure}

We take the multimode interference (MMI) coupler efficiency to be $\eta_\text{MMI}<0.445$, where the equality holds for the value determined from simulations. Figure~\ref{SI_Fig:MMI_PL} shows ${\mu}$PL spectra produced by one of the fabricated devices under $845$~nm continuous wave (CW) laser pumping, collected individually from each of the MMI output ports. Identical spectral features and comparable photon counts at the two ports suggest that the designed MMI $50:50$ split ratio is within reach. 

The grating filter efficiency $\eta_\text{filter}$ was estimated to be $\approx 45~\%$ based on QD spectra taken before and after the filtering. The lensed fiber insertion loss was estimated by injecting CW laser light into it by way of the 10~\% port of a 90:10 fiber beam splitter and measuring the reflected power, after positioning the lensed fiber below the lower SiO$_2$ cladding on the chip. In this case, light launched into free-space by the lensed fiber was reflected at the polished Si substrate facet and coupled back into the former. Assuming only Fresnel reflection at the air-to-Si interface, and perfect coupling back into the lensed fiber, $\eta_\text{LF}\approx0.5$. It is likely, however, that the fiber coupling efficiency for the reflected free-space light was not perfect, so this estimate is at best a lower bound. For an upper bound, we take simply the 0.25~dB insertion loss that is specified for the FC/APC mating sleeve. 

Facet losses~$\eta_\text{facet}$ were estimated from insertion losses measured from three $\approx$ 20~mm long straight ULLWs fabricated on the same chip as the Archimedean spirals described in the main text. In such short straight guides, propagation losses amount to $<3$~dB/m$ \times $0.02~m $<0.1$~dB, so we estimate $\eta_\text{facet}$ as half the insertion loss. From all three waveguides, we obtain $\eta_\text{facet}=0.55\pm0.02$, where the uncertainty corresponds to the standard deviation among all measured values. The fiber that linked the output of the grating filter to the input of the SNSPD was measured to have a transmission of $\approx90~\%$. Finally, we estimated the SNSPD efficiency $\eta_\text{SNSPD}=0.71\pm0.03$ by noting the count rates obtained from a calibrated laser signal at the QD wavelength. The uncertainty here was propagated from power and count rate measurement uncertainties due to experimental fluctuations. 

For a detected SNSPD count rate of $\approx 2.1\times10^5\text{s}^{-1}$, we find, using the MMI efficiency upper bound $\eta_\text{MMI}=0.445$, the QD-ULLW coupling $\eta_\text{QD-ULLW}=R_\text{det.}/80$~MHz to be within a 4~\% to 7~\% interval. The uncertainty here is largely due to the lensed fiber insertion loss $\eta_\text{LF}$. Since however we expect $\eta_\text{MMI}\leq0.445$, then $\eta_\text{QD-ULLW}\geq 4~\%$, conservatively.

As shown in Section~\ref{SI_section:FDTD_sim} of the SI, the QD dipole moment orientation and position have the largest impact on the overall coupling efficiency $\eta_\text{QD-ULLW}$. Since neither position nor orientation of the accessed QD were known or controlled, it is likely that these two factors account for the relatively low observed $\eta_\text{QD-ULLW}$.

\section{Coupling efficiency simulations}
\label{SI_section:FDTD_sim}
To assess the expected QD-to-ULLW coupling efficiency $\eta_\text{QD-ULLW}$, we performed FDTD simulations of an electric point dipole source radiating inside of a hybrid waveguide geometry that approximates that of the tested device. In particular, we attempt to include geometrical imperfections that were apparent from the scanning electron micrograph in Fig.~2(b) of the main text, as well as the QD location and dipole moment orientation with respect to the GaAs host nanowaveguide. Table~\ref{table:eta_FP} shows the coupling efficiency $\eta_\text{QD-ULLW}$ and Purcell factors $F_p$ calculated considering combined variations of $\theta_\text{dip.}$, the dipole orientation with respect to the GaAs waveguide axis; $\Delta x_\text{dip.}$, the dipole displacement from the GaAs WG center; $\Delta x_\text{WG}$, the lateral displacement between the GaAs and $\SiN$ WGs; and $\Delta \theta$, the angular misalignment between the $\SiN$ and GaAs waveguides. The first row on the Table corresponds to the ideal case.
\begin{table}[!h]
\centering
\caption{Simulated dipole quantum efficiency $\eta_\text{QD-ULLW}$ and Purcell factor $F_p$ into the ultra-low loss waveguide for various geometrical parameters defined in the text.}
\label{table:eta_FP}
\begin{tabular}{|c|c|c|c|c|c|}
\hline
\textbf{$\theta_\text{dip.} (^\circ)$} & \textbf{$\Delta x_\text{dip.}$ (nm)} & \textbf{$\Delta x_\text{WG}$ (nm)} & \textbf{$\Delta\theta (^\circ)$} & \textbf{$F_p$} & \textbf{$\eta_\text{QD-ULLW}$} \\ \hline
90                                   & 0                                            & 0                               & 0                                & 2.01     & 0.31                     \\ \hline
90                                   & 0                                            & 500                          & 0                                & 1.96     & 0.22                     \\ \hline
90                                   & 67.5                                      & 0                               & 0                                & 1.27     & 0.28                        \\ \hline
90                                   & 0                                            & 340                          & 0.9                             & 2.06     & 0.30                     \\ \hline
90                                   & 67.5                                      & 340                          & 0.9                             & 1.26     & 0.27                     \\ \hline
0                                    & 0                                            & 0                               & 0                                & 0.87    & 0                     \\ \hline
0                                    & 67.5                                      & 0                               & 0                                & 0.52    & 0.001                   \\ \hline
70                                   & 0                                            & 0                               & 0                                & 1.02     & 0.073                    \\ \hline
\end{tabular}
\end{table}

It is apparent in these results that dipole orientation has the strongest influence in the coupling efficiency, primarily because longitudinally oriented dipoles ($\theta_\text{dip.}=0$ with respect to the GaAs waveguide) does not couple to the (single) TE mode of the GaAs guide. Other geometrical imperfections contribute considerably less, within $\approx$ 30~\%, to a decreased overall coupling efficiency. The Purcell factor $F_p$ is affected almost exclusively by the position and orientation of the quantum dot within the GaAs nanowaveguide. Taken altogether, these results indicate that sub-optimal position and orientation of the (non-deterministically positioned) quantum dot are likely the main contributors to the low observed coupling efficiency. 

\section{Optimized adiabatic taper design}
\label{SI_Section_adiabatic}
\begin{figure}[h!]
    \centering
    \includegraphics[width=0.7\columnwidth]{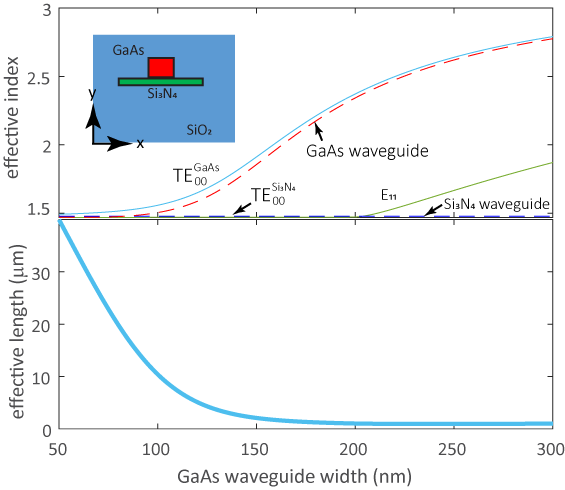}
    \caption{Top: Effective indices of the supermodes of the hybrid waveguide geometry shown in
    the inset, as a function of GaAs waveguide width. Dashed lines are effective indices for the GaAs and $\SiN$ waveguides taken alone. The GaAs and $\SiN$ thicknesses were 190~nm and 40~nm respectively, and the $\SiN$ width was 2~$\mum$. The first two supermodes TE$_\text{00}^\text{GaAs}$ and TE$_\text{00}^{\SiN}$ are considered for the adiabatic taper calculation. A second-order hybrid mode, labeled E$_\text{11}$, appears at a GaAs width of $\approx$200~nm. Bottom: Effective adiabatic coupling length $L_\text{eff}$ for the TE$_\text{00}^\text{GaAs}$ and TE$_\text{00}^{\SiN}$ modes, as a function of GaAs waveguide width.} 
    \label{SI_Fig:adiabatic_taper_neff}
\end{figure}

Although our fabricated devices featured adiabatic couplers with relatively low ($\approx30~\%$) efficiency, below we assert the potential of our device platform regarding efficient single-photon emission into the ULLWs. We do so by designing a mode converter with $> 93$~\% efficiency, following the procedure outlined in ref.~\cite{sun_adiabaticity_2009}, which defines an adiabaticity criterion based on a desired level of coupling loss for a minimized length. The latter characteristic is highly important to ensure stability of free-standing GaAs devices after removal of the AlGaAs sacrificial layer. 
We start by calculating the effective coupling length $L_\text{eff}=\lambda/(n_1-n_2)$ between the first two supermodes of the hybrid waveguide shown in the inset of Fig.~\ref{SI_Fig:adiabatic_taper_neff}, as a function of the GaAs ridge width, $w_\text{GaAs}$. Here, $\lambda=920$~nm is the design wavelength, and $n_1$ and $n_2$ are the supermode effective indices of the TE$_\text{00}^\text{GaAs}$ and TE$_\text{00}^{\SiN}$ modes indicated in the Figure. The effective length plotted in Fig.~\ref{SI_Fig:adiabatic_taper_neff}  varies considerably with $w_\text{GaAs}$, and indicates the necessary length scale for achieving adiabaticity in transitioning from the hybrid to the ULLW.

To determine the GaAs taper width profile $w_\text{GaAs}(z)$ that minimizes coupling to the second-order supermode in the shortest length, we start from the equation~\cite{sun_adiabaticity_2009}
\begin{equation}
    \frac{1}{2\kappa(1+\gamma^2)^{3/2}}\frac{d\gamma}{dz}\leq \sqrt{\epsilon},
\end{equation}
where $\epsilon$ is the (desired) power fraction that is coupled into the (unwanted) second supermode. In this expression, $\gamma=\delta/\kappa$, with $\delta=2\pi(n_\text{eff}^\text{GaAs}-n_\text{eff}^{\SiN})/\lambda$, is the propagation constant mismatch for the two individual, uncoupled waveguides, and $\kappa^2=(S^2-\delta^2)$, with $S=2\pi(n_1-n_2)/\lambda$, is the coupling strength between the two. Both parameters $\gamma$ and $\kappa$ are calculated as a function of $w_\text{GaAs}$, so that we can write 
\begin{equation}
\label{Eq:adiabatic}
    \frac{1}{2\kappa(1+\gamma^2)^{3/2}}\frac{d\gamma}{dw_\text{GaAs}}\leq \sqrt{\epsilon}\frac{dz}{dw_\text{GaAs}}.
\end{equation}

\begin{figure}[h!]
    \centering
    \includegraphics[width=0.7\columnwidth]{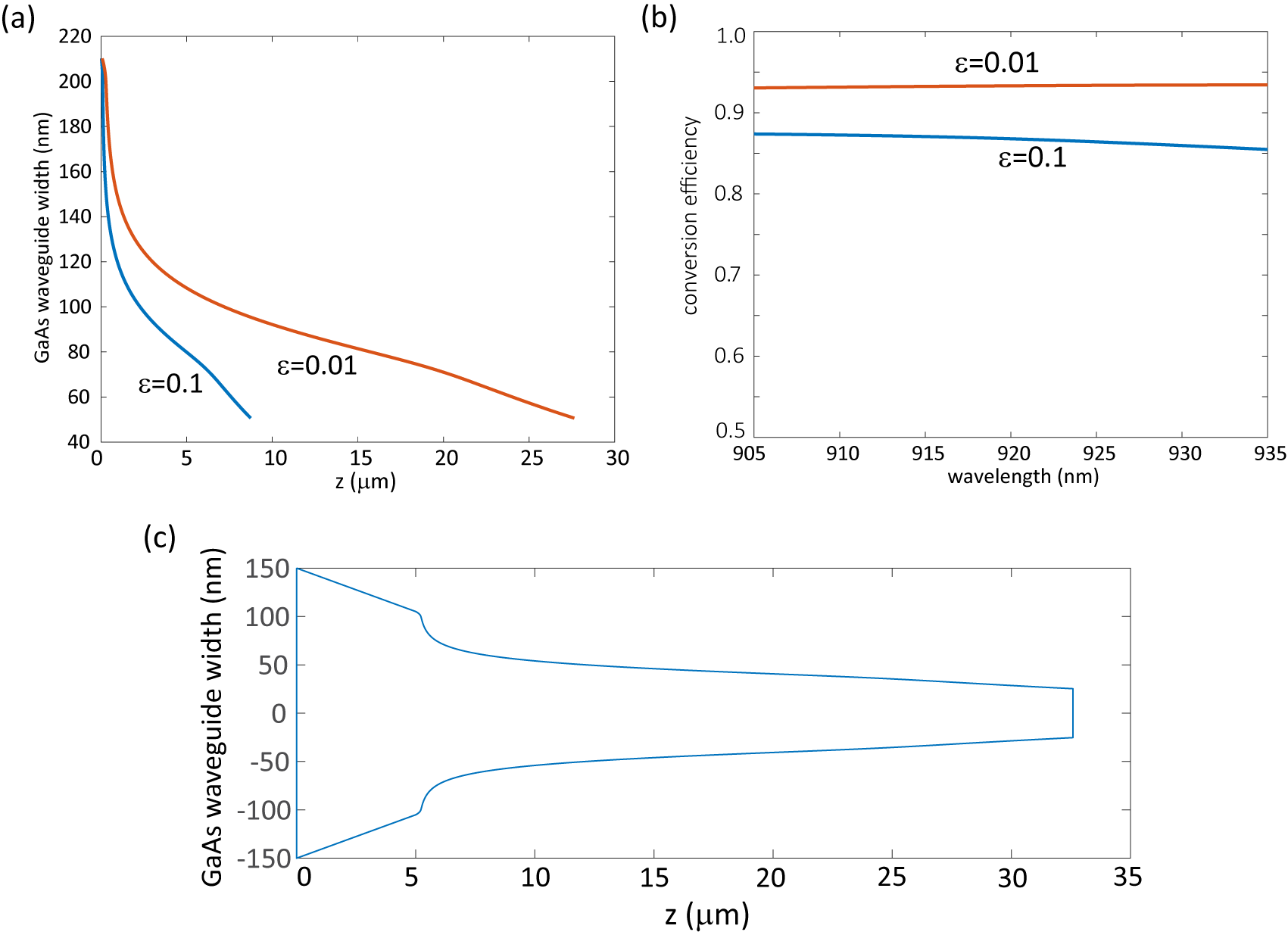}
    \caption{(a) Optimized adiabatic GaAs width taper profiles for $\epsilon=0.1$ and $\epsilon=0.01$. (b) Simulated conversion efficiencies, from the TE$_\text{00}^\text{GaAs}$ mode to the fundamental TE $\SiN$ waveguide mode as a function of wavelength. (c) GaAs waveguide width profile obtained for $\epsilon=0.01$, used to obtain the simulated result in (b).}
    \label{SI_Fig:adiabatic_taper}
\end{figure}

Integrating eq.~(\ref{Eq:adiabatic}) over a sufficiently small width variation $\delta w_\text{GaAs}$ that the first factor on the left hand-side can be assumed to be constant, we obtain the local adiabatic coupling length $\delta z(w_\text{GaAs})$
\begin{equation}
    \delta z(w_\text{GaAs})= \frac{1}{2\sqrt{\epsilon}\kappa(1+\gamma^2)^{3/2}}\delta\gamma(w_\text{GaAs}),
\end{equation}
which we then numerically integrate to obtain an optimized GaAs width profile. Optimized width profiles for our geometry starting from $w_\text{GaAs}=200$~nm are shown in Fig.~\ref{SI_Fig:adiabatic_taper}(a), for $\epsilon=0.1$ and $\epsilon=0.01$. The conversion efficiencies for the two profiles, as a function of wavelength, are shown in Fig.~\ref{SI_Fig:adiabatic_taper},  calculated in finite difference time domain simulations in which the GaAs waveguide is excited with its fundamental TE mode at 920~nm.  Figure ~\ref{SI_Fig:adiabatic_taper}(c) shows the adiabatic taper profile used in the simulation of the $\epsilon=0.01$ case.

\clearpage
\section{Fits for resonance fluorescence spectrum}
\label{SI_Section_RF_fits}
\begin{figure}[h]
    \centering
    \includegraphics[width=1.0\columnwidth]{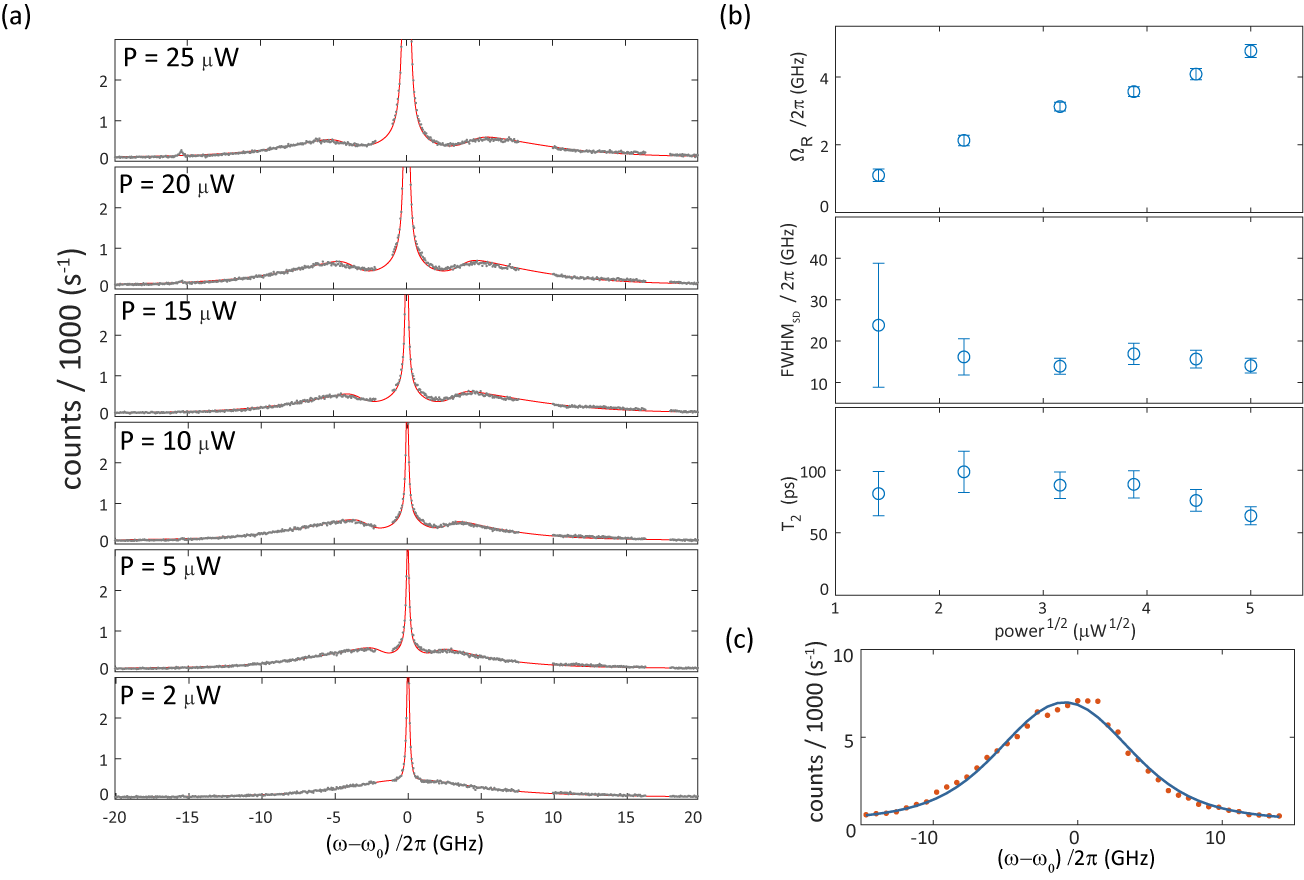}
    \caption{(a) High-resolution resonance fluorescence spectra obtained with a scanning Fabry-Perot resonator (grey dots) and fit (red lines) using the Mollow triplet model of ref.~\cite{konthasinghe_coherent_2012}, for varying nominal excitation power P. (b) Fit parameters as a function of the square-root of the excitation power. $\Omega_R$: Rabi Frequency; FWHM$_\text{SD}$: Gaussian spectral diffusion FWHM; $T_2$: coherence time. Uncertainties are 95~\% fit confidence intervals, corresponding to two standard deviations. (c) Resonance fluorescence intensity as a function of laser detuning from the QD transition, corresponding to excitation laser power of $ 10~\mu$W, extracted from low-resolution spectra from a grating spectrometer, as described in the text. Red dots: experimental data; continuous line:Gaussian fit.}
    \label{SI_Fig:RF_FWHM}
\end{figure}

As described in the main text, the measured Mollow triplet spectra, shown in Fig.~\ref{fig:RF}c in the main text, presented slight asymmetries that suggested laser detuning~\cite{Ulhaq2013} and spectral diffusion~\cite{konthasinghe_coherent_2012}. To obtain a better understanding of such features, we fit the data with the physical model derived in Konthasinghe \emph{et al.}~\cite{konthasinghe_coherent_2012}, which extends Mollow's original results for the case of nonzero laser detuning from the transition. The model in addition assumes that the transition spectrally wanders at time-scales longer than the radiative decay time $T_1$, corresponding to a time-varying detuning which is imprinted in the spectrum via the convolution 
\begin{equation}
I(\nu,\Delta\omega)\propto\int\tilde{g}(\nu,\Delta\omega)\exp\left(-\frac{\Delta\omega^2}{2\sigma_\text{SD}^2}\right)d\Delta\omega.
\end{equation}
Here, $I(\nu,\Delta\omega)$ is the final spectrum, $\tilde{g}(\nu,\Delta\omega)$ is the detuning-dependent Mollow triplet spectrum~\cite{Ulhaq2013,konthasinghe_coherent_2012}, $\Delta\omega$ is the laser-transition detuning, and $\sigma_{SD}$ is a measure of the extent of the spectral diffusion, which is postulated to follow a Gaussian distribution. Figure~\ref{SI_Fig:RF_FWHM}(a) shows fits to the spectra obtained with the parameters in Fig.~\ref{SI_Fig:RF_FWHM}(b), for the corresponding excitation powers $P$. As expected the Rabi frequencies $\Omega_\text{R}$ increase linearly with $\sqrt{P}$. The coherence time is seen to vary somewhat for varying powers, though remaining below 100~ps throughout. The spectral diffusion distribution full-width at half maximum, FWHM$_\text{SD}$, is seen to vary between 15~GHz and 20~GHz, except for the lower power, where the confidence interval is large. To understand whether such a large spectral diffusion is reasonable, we show, in Fig.~\ref{SI_Fig:RF_FWHM}(c) a graph of the resonance fluorescence intensity as a function of laser detuning from the $X_1$ QD transition. In this measurement, low-resolution resonance fluorescence spectra were recorded with a grating spectrometer, and a wavelength meter with a 600~MHz nominal accuracy was used to record the laser frequency. As described in \textbf{Methods}, an above-GaAs-band co-pump laser was used to optically gate the resonance fluorescence emission~\cite{nguyen_optically_2012}. The plotted data correspond to the difference in integrated counts obtained with the co-pump on and off. The experimental data obtained in this fashion (red dots in Fig.~\ref{SI_Fig:RF_FWHM}(c)) could be fit with a Gaussian with an FWHM~$\approx11$~GHz, which is comparable to spectral diffusion FWHMs obtained from the Mollow triplet fits.

\clearpage
\section{Modelling of interferometric correlation spectroscopy}
\label{SI_Section_MZI}
\label{SI_section:RF_PulseG2}
We used a variable delay Mach-Zehnder interferometer to measure the correlation,  $g^{(1)}$,  of the filtered QD emission. To model the visibility of the interference fringes for QD emission under resonant excitation, we used the expression for the first-order correlation function~\cite{proux_measuring_2015}:

\begin{multline}
\label{Eq:SI_MZI_visibility}
    g^{(1)}(\tau) = e^{-\omega_L|\tau|}. \frac{1}{2}e^{-\frac{|\tau|}{T_L}} . \Bigg[\frac{\frac{T_2}{2T_1}}{1+\Omega_\text{R}^2T_1T_2}+  \frac{E}{2}e^{-\frac{|\tau|}{T_2}}+Ge^{-\frac{1}{2}{\big( \frac{|\tau|}{T_2}\big)}^2}+\\ \big(\frac{E}{2}e^{\eta\tau} +Ge^{-\frac{1}{2}{\eta\tau}\big)^2}\big).\big[\alpha \cos(\nu\tau)+\beta \sin(\nu\tau)\big]\Bigg] 
\end{multline}

Here,
\begin{align*}
\eta &=\frac{1}{2}\bigg[\frac{1}{T_1}+\frac{1}{T_2}\bigg] & \nu=&\sqrt{\Omega_\text{R}^2+\frac{1}{4}\bigg[\frac{1}{T_1}-\frac{1}{T_2}}\bigg]^2 \\
\alpha&=1-\frac{T_2}{T_1(1+\Omega_\text{R}^2T_1T_2)} &\beta&=\frac{\Omega_\text{R}^2T_1(3T_2-T_1)-\frac{{(T_1-T_2)}^2}{T_1T_2}}{2\nu T_1(1+\Omega_\text{R}^2T_1T_2)}
\end{align*}

The parameter $T_L$ represents the coherence time of laser, which was estimated to be $20~\mu$s based on the specified laser linewidth, and  $\Omega_\text{R}$ is the Rabi frequency. The visibility traces were fit to estimate $T_2$ and $\Omega_R$, while the measured value of $T_1 = (0.63 \pm 0.01)$~ns was used. 

In eq.~(\ref{Eq:SI_MZI_visibility}), the first term corresponds to a coherent laser background due to incomplete pump suppression which captures interference between laser fields and QD emission fields at resonant excitation. The term in parenthesis corresponds to exponential decay associated with coherence lifetime and oscillatory terms corresponding to the system undergoing Rabi cycles\cite{muller}. In addition, we consider coefficients of the two terms as a statistical sum of exponential and Gaussian  monotonous decay, given by coefficients $E$ and $G$, respectively, accounting for spectral diffusion of the QD.

\begin{figure}[h]
    \centering
    \includegraphics[width=0.5\columnwidth]{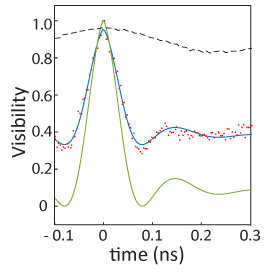}
    \caption{ Mach-Zehnder interferogram visibility (red points) of QD emission when excited resonantly and the corresponding fit (blue curve), normalized to the system response measured with laser as the MZI input (dashed black curve). The offset in the fit model accounts for the resonant laser that is not sufficiently suppressed. The green curve shows the normalized MZI visibility for the same parameter derived from the fit but without the offset.}
    \label{SI_fig:SI_Fig_MZI_FitComponents.png}
\end{figure}

Figure~\ref{SI_fig:SI_Fig_MZI_FitComponents.png} shows the visibility plot (black) for CW laser which was used to characterize the setup response accounting for imperfections in alignment and overlay of the two arms of the MZI. The curve was used as a reference to normalize the visibility traces for QD emission. Experimental data (red dots) and the corresponding fits to the model (blue) shows a representative trace for 
$g^{(1)}$ for resonance fluorescence in the strong drive regime, where Rabi oscillations are observed. The non-zero visibility obtained from fringe data at longer time-delay ($> 0.3$~ns) originates from  laser background. The trace in green shows the fitted function without the laser background. 


\begin{thebibliography}{10}
\expandafter\ifx\csname url\endcsname\relax
  \def\url#1{\texttt{#1}}\fi
\expandafter\ifx\csname urlprefix\endcsname\relax\def\urlprefix{URL }\fi
\providecommand{\bibinfo}[2]{#2}
\providecommand{\eprint}[2][]{\url{#2}}

\bibitem{liu_720_2021}
\bibinfo{author}{Liu, K.} \emph{et~al.}
\newblock \bibinfo{title}{720 {Million} {Quality} {Factor} {Integrated}
  {All}-{Waveguide} {Photonic} {Resonator}}.
\newblock In \emph{\bibinfo{booktitle}{2021 {Device} {Research} {Conference}
  ({DRC})}}, \bibinfo{pages}{1--2} (\bibinfo{year}{2021}).
\newblock \bibinfo{note}{ISSN: 2640-6853}.

\bibitem{blumenthal_photonic_2020}
\bibinfo{author}{Blumenthal, D.~J.}
\newblock \bibinfo{title}{Photonic integration for {UV} to {IR} applications}.
\newblock \emph{\bibinfo{journal}{APL Photonics}} \textbf{\bibinfo{volume}{5}},
  \bibinfo{pages}{020903} (\bibinfo{year}{2020}).
\newblock \urlprefix\url{https://aip.scitation.org/doi/10.1063/1.5131683}.
\newblock \bibinfo{note}{Publisher: American Institute of Physics}.

\bibitem{doerr_silicon_2018}
\bibinfo{author}{Doerr, C.} \& \bibinfo{author}{Chen, L.}
\newblock \bibinfo{title}{Silicon {Photonics} in {Optical} {Coherent}
  {Systems}}.
\newblock \emph{\bibinfo{journal}{Proceedings of the IEEE}}
  \textbf{\bibinfo{volume}{106}}, \bibinfo{pages}{2291--2301}
  (\bibinfo{year}{2018}).
\newblock \bibinfo{note}{Conference Name: Proceedings of the IEEE}.

\bibitem{marpaung_integrated_2013}
\bibinfo{author}{Marpaung, D.} \emph{et~al.}
\newblock \bibinfo{title}{Integrated microwave photonics}.
\newblock \emph{\bibinfo{journal}{Laser \& Photonics Reviews}}
  \textbf{\bibinfo{volume}{7}}, \bibinfo{pages}{506--538}
  (\bibinfo{year}{2013}).
\newblock
  \urlprefix\url{https://onlinelibrary.wiley.com/doi/abs/10.1002/lpor.201200032}.
\newblock \bibinfo{note}{\_eprint:
  https://onlinelibrary.wiley.com/doi/pdf/10.1002/lpor.201200032}.

\bibitem{lai_earth_2020}
\bibinfo{author}{Lai, Y.-H.} \emph{et~al.}
\newblock \bibinfo{title}{Earth rotation measured by a chip-scale ring laser
  gyroscope}.
\newblock \emph{\bibinfo{journal}{Nature Photonics}}
  \textbf{\bibinfo{volume}{14}}, \bibinfo{pages}{345--349}
  (\bibinfo{year}{2020}).
\newblock \urlprefix\url{https://www.nature.com/articles/s41566-020-0588-y}.

\bibitem{newman_architecture_2019}
\bibinfo{author}{Newman, Z.~L.} \emph{et~al.}
\newblock \bibinfo{title}{Architecture for the photonic integration of an
  optical atomic clock}.
\newblock \emph{\bibinfo{journal}{Optica}} \textbf{\bibinfo{volume}{6}},
  \bibinfo{pages}{680--685} (\bibinfo{year}{2019}).
\newblock
  \urlprefix\url{https://www.osapublishing.org/optica/abstract.cfm?uri=optica-6-5-680}.
\newblock \bibinfo{note}{Publisher: Optical Society of America}.

\bibitem{sparrow}
\bibinfo{author}{Sparrow, C.} \emph{et~al.}
\newblock \bibinfo{title}{Simulating the vibrational quantum dynamics of
  molecules using photonics}.
\newblock \emph{\bibinfo{journal}{Nature}} \textbf{\bibinfo{volume}{557}},
  \bibinfo{pages}{660--667} (\bibinfo{year}{2018}).
\newblock \urlprefix\url{https://doi.org/10.1038/s41586-018-0152-9}.

\bibitem{steinbrecher2018quantum}
\bibinfo{author}{Steinbrecher, G.~R.}, \bibinfo{author}{Olson, J.~P.},
  \bibinfo{author}{Englund, D.} \& \bibinfo{author}{Carolan, J.}
\newblock \bibinfo{title}{Quantum optical neural networks}.
\newblock \emph{\bibinfo{journal}{npj Quantum Information}}
  \textbf{\bibinfo{volume}{5}}, \bibinfo{pages}{60} (\bibinfo{year}{2019}).

\bibitem{choi2019percolation}
\bibinfo{author}{Choi, H.}, \bibinfo{author}{Pant, M.}, \bibinfo{author}{Guha,
  S.} \& \bibinfo{author}{Englund, D.}
\newblock \bibinfo{title}{Percolation-based architecture for cluster state
  creation using photon-mediated entanglement between atomic memories}.
\newblock \emph{\bibinfo{journal}{npj Quantum Information}}
  \textbf{\bibinfo{volume}{5}}, \bibinfo{pages}{1--7} (\bibinfo{year}{2019}).

\bibitem{Rudolph:2017du}
\bibinfo{author}{Rudolph, T.}
\newblock \bibinfo{title}{{Why I am optimistic about the silicon-photonic route
  to quantum computing}}.
\newblock \emph{\bibinfo{journal}{APL Photonics}} \textbf{\bibinfo{volume}{2}},
  \bibinfo{pages}{030901--20} (\bibinfo{year}{2017}).

\bibitem{Wang2019}
\bibinfo{author}{Wang, H.} \emph{et~al.}
\newblock \bibinfo{title}{Boson sampling with 20 input photons and a 60-mode
  interferometer in a $1{0}^{14}$-dimensional hilbert space}.
\newblock \emph{\bibinfo{journal}{Physical Review Letters}}
  \textbf{\bibinfo{volume}{123}}, \bibinfo{pages}{250503}
  (\bibinfo{year}{2019}).
\newblock
  \urlprefix\url{https://link.aps.org/doi/10.1103/PhysRevLett.123.250503}.

\bibitem{gimeno-segovia_relative_2017}
\bibinfo{author}{Gimeno-Segovia, M.} \emph{et~al.}
\newblock \bibinfo{title}{Relative multiplexing for minimising switching in
  linear-optical quantum computing}.
\newblock \emph{\bibinfo{journal}{New Journal of Physics}}
  \textbf{\bibinfo{volume}{19}}, \bibinfo{pages}{063013}
  (\bibinfo{year}{2017}).
\newblock \urlprefix\url{https://doi.org/10.1088/1367-2630/aa7095}.
\newblock \bibinfo{note}{Publisher: IOP Publishing}.

\bibitem{brod_photonic_2019}
\bibinfo{author}{Brod, D.~J.} \emph{et~al.}
\newblock \bibinfo{title}{Photonic implementation of boson sampling: a review}.
\newblock \emph{\bibinfo{journal}{Advanced Photonics}}
  \textbf{\bibinfo{volume}{1}}, \bibinfo{pages}{034001} (\bibinfo{year}{2019}).
\newblock
  \urlprefix\url{https://www.spiedigitallibrary.org/journals/advanced-photonics/volume-1/issue-3/034001/Photonic-implementation-of-boson-sampling-a-review/10.1117/1.AP.1.3.034001.full}.
\newblock \bibinfo{note}{Publisher: SPIE}.

\bibitem{deshpande_quantum_2022}
\bibinfo{author}{Deshpande, A.} \emph{et~al.}
\newblock \bibinfo{title}{Quantum computational advantage via high-dimensional
  {Gaussian} boson sampling}.
\newblock \emph{\bibinfo{journal}{Science Advances}}  (\bibinfo{year}{2022}).
\newblock
  \urlprefix\url{https://www.science.org/doi/abs/10.1126/sciadv.abi7894}.

\bibitem{Eisaman2011}
\bibinfo{author}{Eisaman, M.~D.}, \bibinfo{author}{Fan, J.},
  \bibinfo{author}{Migdall, A.} \& \bibinfo{author}{Polyakov, S.~V.}
\newblock \bibinfo{title}{Invited review article: Single-photon sources and
  detectors}.
\newblock \emph{\bibinfo{journal}{Review of Scientific Instruments}}
  \textbf{\bibinfo{volume}{82}}, \bibinfo{pages}{071101}
  (\bibinfo{year}{2011}).
\newblock \urlprefix\url{http://link.aip.org/link/?RSI/82/071101/1}.

\bibitem{kaneda_high-efficiency_nodate}
\bibinfo{author}{Kaneda, F.} \& \bibinfo{author}{Kwiat, P.~G.}
\newblock \bibinfo{title}{High-efficiency single-photon generation via
  large-scale active time multiplexing}.
\newblock \emph{\bibinfo{journal}{Science Advances}}
  \textbf{\bibinfo{volume}{5}}, \bibinfo{pages}{eaaw8586}
  (\bibinfo{year}{2021}).
\newblock \urlprefix\url{https://www.science.org/doi/10.1126/sciadv.aaw8586}.
\newblock \bibinfo{note}{Publisher: American Association for the Advancement of
  Science}.

\bibitem{ji_methods_2021}
\bibinfo{author}{Ji, X.}, \bibinfo{author}{Roberts, S.},
  \bibinfo{author}{Corato-Zanarella, M.} \& \bibinfo{author}{Lipson, M.}
\newblock \bibinfo{title}{Methods to achieve ultra-high quality factor silicon
  nitride resonators}.
\newblock \emph{\bibinfo{journal}{APL Photonics}} \textbf{\bibinfo{volume}{6}},
  \bibinfo{pages}{071101} (\bibinfo{year}{2021}).
\newblock \urlprefix\url{https://aip.scitation.org/doi/full/10.1063/5.0057881}.
\newblock \bibinfo{note}{Publisher: American Institute of Physics}.

\bibitem{Somaschi2016}
\bibinfo{author}{Somaschi, N.} \emph{et~al.}
\newblock \bibinfo{title}{Near-optimal single-photon sources in the solid
  state}.
\newblock \emph{\bibinfo{journal}{Nature Photonics}}
  \textbf{\bibinfo{volume}{10}}, \bibinfo{pages}{340--345}
  (\bibinfo{year}{2016}).
\newblock \urlprefix\url{https://doi.org/10.1038/nphoton.2016.23}.

\bibitem{Dietrich2016}
\bibinfo{author}{Dietrich, C.~P.}, \bibinfo{author}{Fiore, A.},
  \bibinfo{author}{Thompson, M.~G.}, \bibinfo{author}{Kamp, M.} \&
  \bibinfo{author}{H\"ofling, S.}
\newblock \bibinfo{title}{GaAs integrated quantum photonics: Towards compact
  and multi-functional quantum photonic integrated circuits}.
\newblock \emph{\bibinfo{journal}{Laser Photonics Rev.}}
  \textbf{\bibinfo{volume}{10}}, \bibinfo{pages}{870--894}
  (\bibinfo{year}{2016}).
\newblock \urlprefix\url{http://dx.doi.org/10.1002/lpor.201500321}.

\bibitem{Lodahl2018}
\bibinfo{author}{Lodahl, P.}
\newblock \bibinfo{title}{Quantum-dot based photonic quantum networks}.
\newblock \emph{\bibinfo{journal}{Quantum Science and Technology}}
  \textbf{\bibinfo{volume}{3}}, \bibinfo{pages}{013001} (\bibinfo{year}{2018}).
\newblock \urlprefix\url{http://stacks.iop.org/2058-9565/3/i=1/a=013001}.

\bibitem{elshaari_hybrid_2020}
\bibinfo{author}{Elshaari, A.~W.}, \bibinfo{author}{Pernice, W.},
  \bibinfo{author}{Srinivasan, K.}, \bibinfo{author}{Benson, O.} \&
  \bibinfo{author}{Zwiller, V.}
\newblock \bibinfo{title}{Hybrid integrated quantum photonic circuits}.
\newblock \emph{\bibinfo{journal}{Nat. Photonics}}
  \textbf{\bibinfo{volume}{14}}, \bibinfo{pages}{285--298}
  (\bibinfo{year}{2020}).
\newblock \urlprefix\url{http://www.nature.com/articles/s41566-020-0609-x}.

\bibitem{kim_hybrid_2020}
\bibinfo{author}{Kim, J.-H.}, \bibinfo{author}{Aghaeimeibodi, S.},
  \bibinfo{author}{Carolan, J.}, \bibinfo{author}{Englund, D.} \&
  \bibinfo{author}{Waks, E.}
\newblock \bibinfo{title}{Hybrid integration methods for on-chip quantum
  photonics}.
\newblock \emph{\bibinfo{journal}{Optica, {OPTICA}}}
  \textbf{\bibinfo{volume}{7}}, \bibinfo{pages}{291--308}
  (\bibinfo{year}{2020}).
\newblock
  \urlprefix\url{https://www.osapublishing.org/optica/abstract.cfm?uri=optica-7-4-291}.

\bibitem{mouradian_scalable_2015}
\bibinfo{author}{Mouradian, S.~L.} \emph{et~al.}
\newblock \bibinfo{title}{Scalable {Integration} of {Long}-{Lived} {Quantum}
  {Memories} into a {Photonic} {Circuit}}.
\newblock \emph{\bibinfo{journal}{Physical Review X}}
  \textbf{\bibinfo{volume}{5}}, \bibinfo{pages}{031009} (\bibinfo{year}{2015}).
\newblock \urlprefix\url{http://link.aps.org/doi/10.1103/PhysRevX.5.031009}.

\bibitem{Davanco2017}
\bibinfo{author}{Davanco, M.} \emph{et~al.}
\newblock \bibinfo{title}{Heterogeneous integration for on-chip quantum
  photonic circuits with single quantum dot devices}.
\newblock \emph{\bibinfo{journal}{Nature Communications}}
  \textbf{\bibinfo{volume}{8}}, \bibinfo{pages}{889} (\bibinfo{year}{2017}).
\newblock \urlprefix\url{https://doi.org/10.1038/s41467-017-00987-6}.

\bibitem{wan_large-scale_2020}
\bibinfo{author}{Wan, N.~H.} \emph{et~al.}
\newblock \bibinfo{title}{Large-scale integration of artificial atoms in hybrid
  photonic circuits}.
\newblock \emph{\bibinfo{journal}{Nature}} \textbf{\bibinfo{volume}{583}},
  \bibinfo{pages}{226--231} (\bibinfo{year}{2020}).
\newblock \urlprefix\url{https://www.nature.com/articles/s41586-020-2441-3}.

\bibitem{eraerds_photon_2010}
\bibinfo{author}{Eraerds, P.}, \bibinfo{author}{Legre, M.},
  \bibinfo{author}{Zhang, J.}, \bibinfo{author}{Zbinden, H.} \&
  \bibinfo{author}{Gisin, N.}
\newblock \bibinfo{title}{Photon counting otdr: Advantages and limitations}.
\newblock \emph{\bibinfo{journal}{Journal of Lightwave Technology}}
  \textbf{\bibinfo{volume}{28}}, \bibinfo{pages}{952--964}
  (\bibinfo{year}{2010}).

\bibitem{flagg_resonantly_2009}
\bibinfo{author}{Flagg, E.~B.} \emph{et~al.}
\newblock \bibinfo{title}{Resonantly driven coherent oscillations in a
  solid-state quantum emitter}.
\newblock \emph{\bibinfo{journal}{Nature Physics}}
  \textbf{\bibinfo{volume}{5}}, \bibinfo{pages}{203--207}
  (\bibinfo{year}{2009}).
\newblock \urlprefix\url{http://www.nature.com/articles/nphys1184}.

\bibitem{nick_vamivakas_spin-resolved_2009}
\bibinfo{author}{Nick~Vamivakas, A.}, \bibinfo{author}{Zhao, Y.},
  \bibinfo{author}{Lu, C.-Y.} \& \bibinfo{author}{Atat{\"u}re, M.}
\newblock \bibinfo{title}{Spin-resolved quantum-dot resonance fluorescence}.
\newblock \emph{\bibinfo{journal}{Nature Physics}}
  \textbf{\bibinfo{volume}{5}}, \bibinfo{pages}{198--202}
  (\bibinfo{year}{2009}).
\newblock \urlprefix\url{https://www.nature.com/articles/nphys1182}.

\bibitem{ulhaq_cascaded_2012}
\bibinfo{author}{Ulhaq, A.} \emph{et~al.}
\newblock \bibinfo{title}{Cascaded single-photon emission from the {Mollow}
  triplet sidebands of a quantum dot}.
\newblock \emph{\bibinfo{journal}{Nature Photonics}}
  \textbf{\bibinfo{volume}{6}}, \bibinfo{pages}{238--242}
  (\bibinfo{year}{2012}).
\newblock \urlprefix\url{https://www.nature.com/articles/nphoton.2012.23}.

\bibitem{lopez_carreno_photon_2017}
\bibinfo{author}{L{\'o}pez~Carreno, J.~C.}, \bibinfo{author}{del Valle, E.} \&
  \bibinfo{author}{Laussy, F.~P.}
\newblock \bibinfo{title}{Photon correlations from the {Mollow} triplet}.
\newblock \emph{\bibinfo{journal}{Laser \& Photonics Reviews}}
  \textbf{\bibinfo{volume}{11}}, \bibinfo{pages}{1700090}
  (\bibinfo{year}{2017}).
\newblock
  \urlprefix\url{https://onlinelibrary.wiley.com/doi/abs/10.1002/lpor.201700090}.

\bibitem{chauhan_ultra-low_2020}
\bibinfo{author}{Chauhan, N.} \emph{et~al.}
\newblock \bibinfo{title}{Ultra-{Low} {Loss} 698 nm and 450 nm visible light
  waveguides for integrated atomic, molecular, and quantum photonics}
  (\bibinfo{year}{2022}).
\newblock \bibinfo{note}{(in press)}.

\bibitem{schnauber2019indistinguishable}
\bibinfo{author}{Schnauber, P.} \emph{et~al.}
\newblock \bibinfo{title}{Indistinguishable photons from deterministically
  integrated single quantum dots in heterogeneous {GaAs/Si$_3$N$_4$} quantum photonic
  circuits}.
\newblock \emph{\bibinfo{journal}{Nano Letters}} \textbf{\bibinfo{volume}{19}},
  \bibinfo{pages}{7164--7172} (\bibinfo{year}{2019}).

\bibitem{katsumi_transfer-printed_2018}
\bibinfo{author}{Katsumi, R.}, \bibinfo{author}{Ota, Y.},
  \bibinfo{author}{Kakuda, M.}, \bibinfo{author}{Iwamoto, S.} \&
  \bibinfo{author}{Arakawa, Y.}
\newblock \bibinfo{title}{Transfer-printed single-photon sources coupled to
  wire waveguides}.
\newblock \emph{\bibinfo{journal}{Optica}} \textbf{\bibinfo{volume}{5}},
  \bibinfo{pages}{691--694} (\bibinfo{year}{2018}).
\newblock
  \urlprefix\url{https://www.osapublishing.org/optica/abstract.cfm?uri=optica-5-6-691}.

\bibitem{Liu2019}
\bibinfo{author}{Liu, J.} \emph{et~al.}
\newblock \bibinfo{title}{A solid-state source of strongly entangled photon
  pairs with high brightness and indistinguishability}.
\newblock \emph{\bibinfo{journal}{Nature Nanotechnology}}
  \textbf{\bibinfo{volume}{14}}, \bibinfo{pages}{586--593}
  (\bibinfo{year}{2019}).
\newblock \urlprefix\url{https://doi.org/10.1038/s41565-019-0435-9}.

\bibitem{bauters}
\bibinfo{author}{Bauters, J.~F.} \emph{et~al.}
\newblock \bibinfo{title}{Planar waveguides with less than 0.1 {dB/m} propagation
  loss fabricated with wafer bonding}.
\newblock \emph{\bibinfo{journal}{Optics Express}}
  \textbf{\bibinfo{volume}{19}}, \bibinfo{pages}{24090--24101}
  (\bibinfo{year}{2011}).
\newblock
  \urlprefix\url{http://www.opticsexpress.org/abstract.cfm?URI=oe-19-24-24090}.

\bibitem{Kuhlmann2013a}
\bibinfo{author}{Kuhlmann, A.~V.} \emph{et~al.}
\newblock \bibinfo{title}{Charge noise and spin noise in a semiconductor
  quantum device}.
\newblock \emph{\bibinfo{journal}{Nat Phys}} \textbf{\bibinfo{volume}{9}},
  \bibinfo{pages}{570--575} (\bibinfo{year}{2013}).
\newblock \urlprefix\url{http://dx.doi.org/10.1038/nphys2688}.

\bibitem{wang_towards_2019}
\bibinfo{author}{Wang, H.} \emph{et~al.}
\newblock \bibinfo{title}{Towards optimal single-photon sources from polarized
  microcavities}.
\newblock \emph{\bibinfo{journal}{Nature Photonics}}
  \textbf{\bibinfo{volume}{13}}, \bibinfo{pages}{770--775}
  (\bibinfo{year}{2019}).
\newblock \urlprefix\url{https://www.nature.com/articles/s41566-019-0494-3}.

\bibitem{ates_post-selected_2009}
\bibinfo{author}{Ates, S.} \emph{et~al.}
\newblock \bibinfo{title}{Post-selected indistinguishable photons from the
  resonance fluorescence of a single quantum dot in a microcavity}.
\newblock \emph{\bibinfo{journal}{Physical Review Letters}}
  \textbf{\bibinfo{volume}{103}}, \bibinfo{pages}{167402}
  (\bibinfo{year}{2009}).
\newblock
  \urlprefix\url{https://link.aps.org/doi/10.1103/PhysRevLett.103.167402}.

\bibitem{huber_filter-free_2020}
\bibinfo{author}{Huber, T.} \emph{et~al.}
\newblock \bibinfo{title}{Filter-free single-photon quantum dot resonance
  fluorescence in an integrated cavity-waveguide device}.
\newblock \emph{\bibinfo{journal}{Optica}} \textbf{\bibinfo{volume}{7}},
  \bibinfo{pages}{380--385} (\bibinfo{year}{2020}).
\newblock
  \urlprefix\url{https://www.osapublishing.org/optica/abstract.cfm?uri=optica-7-5-380}.

\bibitem{he_coherently_2019}
\bibinfo{author}{He, Y.-M.} \emph{et~al.}
\newblock \bibinfo{title}{Coherently driving a single quantum two-level system
  with dichromatic laser pulses}.
\newblock \emph{\bibinfo{journal}{Nature Physics}}
  \textbf{\bibinfo{volume}{15}}, \bibinfo{pages}{941--946}
  (\bibinfo{year}{2019}).

\bibitem{dusanowski_near-unity_2019}
\bibinfo{author}{Dusanowski, L.}, \bibinfo{author}{Kwon, S.-H.},
  \bibinfo{author}{Schneider, C.} \& \bibinfo{author}{H\"ofling, S.}
\newblock \bibinfo{title}{Near-unity indistinguishability single photon source
  for large-scale integrated quantum optics}.
\newblock \emph{\bibinfo{journal}{Physical Review Letters}}
  \textbf{\bibinfo{volume}{122}}, \bibinfo{pages}{173602}
  (\bibinfo{year}{2019}).
\newblock
  \urlprefix\url{https://link.aps.org/doi/10.1103/PhysRevLett.122.173602}.

\bibitem{makhonin}
\bibinfo{author}{Makhonin, M.~N.} \emph{et~al.}
\newblock \bibinfo{title}{Waveguide coupled resonance fluorescence from on-chip
  quantum emitter}.
\newblock \emph{\bibinfo{journal}{Nano Letters}} \textbf{\bibinfo{volume}{14}},
  \bibinfo{pages}{6997--7002} (\bibinfo{year}{2014}).
\newblock \urlprefix\url{https://doi.org/10.1021/nl5032937}.
\newblock \eprint{https://doi.org/10.1021/nl5032937}.

\bibitem{reithmaier_on_chip_NL_2015}
\bibinfo{author}{Reithmaier, G.} \emph{et~al.}
\newblock \bibinfo{title}{On-chip generation, routing, and detection of
  resonance fluorescence}.
\newblock \emph{\bibinfo{journal}{Nano Letters}} \textbf{\bibinfo{volume}{15}},
  \bibinfo{pages}{5208--5213} (\bibinfo{year}{2015}).
\newblock \urlprefix\url{https://doi.org/10.1021/acs.nanolett.5b01444}.
\newblock \eprint{https://doi.org/10.1021/acs.nanolett.5b01444}.

\bibitem{Matthiesen2012}
\bibinfo{author}{Matthiesen, C.}, \bibinfo{author}{Vamivakas, A.~N.} \&
  \bibinfo{author}{Atat\"ure, M.}
\newblock \bibinfo{title}{Subnatural linewidth single photons from a quantum
  dot}.
\newblock \emph{\bibinfo{journal}{Physical Review Letters}}
  \textbf{\bibinfo{volume}{108}}, \bibinfo{pages}{093602}
  (\bibinfo{year}{2012}).
\newblock
  \urlprefix\url{http://link.aps.org/doi/10.1103/PhysRevLett.108.093602}.

\bibitem{konthasinghe_coherent_2012}
\bibinfo{author}{Konthasinghe, K.} \emph{et~al.}
\newblock \bibinfo{title}{Coherent versus incoherent light scattering from a
  quantum dot}.
\newblock \emph{\bibinfo{journal}{Physical Review B}}
  \textbf{\bibinfo{volume}{85}}, \bibinfo{pages}{235315}
  (\bibinfo{year}{2012}).
\newblock \urlprefix\url{https://link.aps.org/doi/10.1103/PhysRevB.85.235315}.
\newblock \bibinfo{note}{Publisher: American Physical Society}.

\bibitem{Thyrrestrup2018}
\bibinfo{author}{Thyrrestrup, H.} \emph{et~al.}
\newblock \bibinfo{title}{Quantum optics with near-lifetime-limited quantum-dot
  transitions in a nanophotonic waveguide}.
\newblock \emph{\bibinfo{journal}{Nano Letters}} \textbf{\bibinfo{volume}{18}},
  \bibinfo{pages}{1801--1806} (\bibinfo{year}{2018}).
\newblock \urlprefix\url{https://doi.org/10.1021/acs.nanolett.7b05016}.
\newblock \eprint{https://doi.org/10.1021/acs.nanolett.7b05016}.

\bibitem{uppu_-chip_2020}
\bibinfo{author}{Uppu, R.} \emph{et~al.}
\newblock \bibinfo{title}{On-chip deterministic operation of quantum dots in
  dual-mode waveguides for a plug-and-play single-photon source}.
\newblock \emph{\bibinfo{journal}{Nature Communications}}
  \textbf{\bibinfo{volume}{11}}, \bibinfo{pages}{3782} (\bibinfo{year}{2020}).
\newblock \urlprefix\url{https://www.nature.com/articles/s41467-020-17603-9}.

\bibitem{errando-herranz_resonance_2021}
\bibinfo{author}{Errando-Herranz, C.} \emph{et~al.}
\newblock \bibinfo{title}{Resonance fluorescence from waveguide-coupled,
  strain-localized, two-dimensional quantum emitters}.
\newblock \emph{\bibinfo{journal}{{ACS} Photonics}}
  \textbf{\bibinfo{volume}{8}}, \bibinfo{pages}{1069--1076}
  (\bibinfo{year}{2021}).
\newblock \urlprefix\url{https://doi.org/10.1021/acsphotonics.0c01653}.

\bibitem{nguyen_optically_2012}
\bibinfo{author}{Nguyen, H.~S.} \emph{et~al.}
\newblock \bibinfo{title}{Optically gated resonant emission of single quantum
  dots}.
\newblock \emph{\bibinfo{journal}{Physical Review Letters}}
  \textbf{\bibinfo{volume}{108}}, \bibinfo{pages}{057401}
  (\bibinfo{year}{2012}).

\bibitem{Davanco2014}
\bibinfo{author}{Davanco, M.}, \bibinfo{author}{Hellberg, C.~S.},
  \bibinfo{author}{Ates, S.}, \bibinfo{author}{Badolato, A.} \&
  \bibinfo{author}{Srinivasan, K.}
\newblock \bibinfo{title}{Multiple time scale blinking in {InAs} quantum dot
  single-photon sources}.
\newblock \emph{\bibinfo{journal}{Physical Review B}}
  \textbf{\bibinfo{volume}{89}} (\bibinfo{year}{2014}).
\newblock \urlprefix\url{http://dx.doi.org/10.1103/PhysRevB.89.161303}.

\bibitem{gazzano_effects_2018}
\bibinfo{author}{Gazzano, O.} \emph{et~al.}
\newblock \bibinfo{title}{Effects of resonant-laser excitation on the emission
  properties in a single quantum dot}.
\newblock \emph{\bibinfo{journal}{Optica}} \textbf{\bibinfo{volume}{5}},
  \bibinfo{pages}{354--359} (\bibinfo{year}{2018}).
\newblock
  \urlprefix\url{https://www.osapublishing.org/optica/abstract.cfm?uri=optica-5-4-354}.

\bibitem{Liu2018}
\bibinfo{author}{Liu, F.} \emph{et~al.}
\newblock \bibinfo{title}{High purcell factor generation of indistinguishable
  on-chip single photons}.
\newblock \emph{\bibinfo{journal}{Nature Nanotechnology}}
  \textbf{\bibinfo{volume}{13}}, \bibinfo{pages}{835--840}
  (\bibinfo{year}{2018}).
\newblock \urlprefix\url{https://doi.org/10.1038/s41565-018-0188-x}.

\bibitem{Ulhaq2013}
\bibinfo{author}{Ulhaq, A.} \emph{et~al.}
\newblock \bibinfo{title}{Detuning-dependent {Mollow} triplet of a
  coherently-driven single quantum dot}.
\newblock \emph{\bibinfo{journal}{Optics Express}}
  \textbf{\bibinfo{volume}{21}}, \bibinfo{pages}{4382--4395}
  (\bibinfo{year}{2013}).
\newblock
  \urlprefix\url{http://www.opticsexpress.org/abstract.cfm?URI=oe-21-4-4382}.

\bibitem{Berthelot2006_NatPhys_motionalnarrowing}
\bibinfo{author}{Berthelot, A.} \emph{et~al.}
\newblock \bibinfo{title}{Unconventional motional narrowing in the optical
  spectrum of a semiconductor quantum dot}.
\newblock \emph{\bibinfo{journal}{Nature Physics}}
  \textbf{\bibinfo{volume}{2}}, \bibinfo{pages}{759--764}
  (\bibinfo{year}{2006}).
\newblock \urlprefix\url{https://doi.org/10.1038/nphys433}.

\bibitem{proux_measuring_2015}
\bibinfo{author}{Proux, R.} \emph{et~al.}
\newblock \bibinfo{title}{Measuring the {Photon} {Coalescence} {Time} {Window}
  in the {Continuous}-{Wave} {Regime} for {Resonantly} {Driven} {Semiconductor}
  {Quantum} {Dots}}.
\newblock \emph{\bibinfo{journal}{Physical Review Letters}}
  \textbf{\bibinfo{volume}{114}}, \bibinfo{pages}{067401}
  (\bibinfo{year}{2015}).
\newblock
  \urlprefix\url{https://link.aps.org/doi/10.1103/PhysRevLett.114.067401}.

\bibitem{Liu2018b}
\bibinfo{author}{Liu, J.} \emph{et~al.}
\newblock \bibinfo{title}{Single self-assembled {InAs/GaAs} quantum dots in
  photonic nanostructures: The role of nanofabrication}.
\newblock \emph{\bibinfo{journal}{Physical Review Applied}}
  \textbf{\bibinfo{volume}{9}}, \bibinfo{pages}{064019} (\bibinfo{year}{2018}).

\bibitem{Pucket2020}
\bibinfo{author}{Puckett, M.~W.} \emph{et~al.}
\newblock \bibinfo{title}{422 million {Q} photonic integrated waveguide
  resonator with a 3.4 billion absorption limit and sub-{MHz} linewidth}.
\newblock \emph{\bibinfo{journal}{Nat. Commun.}} \textbf{\bibinfo{volume}{12}},
  \bibinfo{pages}{934} (\bibinfo{year}{2020}).

\bibitem{Huffman2017}
\bibinfo{author}{Huffman, T.}, \bibinfo{author}{Davenport, M.},
  \bibinfo{author}{Belt, M.}, \bibinfo{author}{Bowers, J.~E.} \&
  \bibinfo{author}{Blumenthal, D.~J.}
\newblock \bibinfo{title}{Ultra-low loss large area waveguide coils for
  integrated optical gyroscopes}.
\newblock \emph{\bibinfo{journal}{IEEE Photonics Technology Letters}}
  \textbf{\bibinfo{volume}{29}}, \bibinfo{pages}{185--188}
  (\bibinfo{year}{2017}).

\bibitem{Huffman2018}
\bibinfo{author}{Huffman, T.~A.} \emph{et~al.}
\newblock \bibinfo{title}{Integrated resonators in an ultralow loss {Si$_3$N$_4$/SiO$_2$}
  platform for multifunction applications}.
\newblock \emph{\bibinfo{journal}{IEEE Journal of Selected Topics in Quantum
  Electronics}} \textbf{\bibinfo{volume}{24}}, \bibinfo{pages}{1--9}
  (\bibinfo{year}{2018}).

\bibitem{moreira_compact_2015}
\bibinfo{author}{Moreira, R.}, \bibinfo{author}{Gundavarapu, S.} \&
  \bibinfo{author}{Blumenthal, D.}
\newblock \bibinfo{title}{Compact programmable monolithically integrated
  10-stage multi-channel {WDM} dispersion equalizer on low-loss silicon nitride
  planar waveguide platform}.
\newblock In \emph{\bibinfo{booktitle}{2015 {Optical} {Fiber} {Communications}
  {Conference} and {Exhibition} ({OFC})}}, \bibinfo{pages}{1--3}
  (\bibinfo{year}{2015}).

\end{thebibliography}

\begin{thebibliography}{10}
\expandafter\ifx\csname url\endcsname\relax
  \def\url#1{\texttt{#1}}\fi
\expandafter\ifx\csname urlprefix\endcsname\relax\def\urlprefix{URL }\fi
\providecommand{\bibinfo}[2]{#2}
\providecommand{\eprint}[2][]{\url{#2}}

\bibitem{ceccarelli_low_2020}
\bibinfo{author}{Ceccarelli, F.} \emph{et~al.}
\newblock \bibinfo{title}{Low {Power} {Reconfigurability} and {Reduced}
  {Crosstalk} in {Integrated} {Photonic} {Circuits} {Fabricated} by
  {Femtosecond} {Laser} {Micromachining}}.
\newblock \emph{\bibinfo{journal}{Laser \& Photonics Reviews}}
  \textbf{\bibinfo{volume}{14}}, \bibinfo{pages}{2000024}
  (\bibinfo{year}{2020}).
\newblock
  \urlprefix\url{https://onlinelibrary.wiley.com/doi/abs/10.1002/lpor.202000024}.
\newblock \bibinfo{note}{\_eprint:
  https://onlinelibrary.wiley.com/doi/pdf/10.1002/lpor.202000024}.

\bibitem{spagnolo_experimental_2014}
\bibinfo{author}{Spagnolo, N.} \emph{et~al.}
\newblock \bibinfo{title}{Experimental validation of photonic boson sampling}.
\newblock \emph{\bibinfo{journal}{Nature Photonics}}
  \textbf{\bibinfo{volume}{8}}, \bibinfo{pages}{615--620}
  (\bibinfo{year}{2014}).
\newblock \urlprefix\url{https://www.nature.com/articles/nphoton.2014.135}.

\bibitem{posner_high-birefringence_2018}
\bibinfo{author}{Posner, M.~T.} \emph{et~al.}
\newblock \bibinfo{title}{High-birefringence direct {UV}-written waveguides for
  use as heralded single-photon sources at telecommunication wavelengths}.
\newblock \emph{\bibinfo{journal}{Optics Express}}
  \textbf{\bibinfo{volume}{26}}, \bibinfo{pages}{24678--24686}
  (\bibinfo{year}{2018}).
\newblock
  \urlprefix\url{https://www.osapublishing.org/oe/abstract.cfm?uri=oe-26-19-24678}.
\newblock \bibinfo{note}{Publisher: Optical Society of America}.

\bibitem{taballione_88_2019}
\bibinfo{author}{Taballione, C.} \emph{et~al.}
\newblock \bibinfo{title}{8x8 reconfigurable quantum photonic processor based
  on silicon nitride waveguides}.
\newblock \emph{\bibinfo{journal}{Optics Express}}
  \textbf{\bibinfo{volume}{27}}, \bibinfo{pages}{26842--26857}
  (\bibinfo{year}{2019}).
\newblock
  \urlprefix\url{https://www.osapublishing.org/oe/abstract.cfm?uri=oe-27-19-26842}.
\newblock \bibinfo{note}{Publisher: Optical Society of America}.

\bibitem{moss_new_2013}
\bibinfo{author}{Moss, D.~J.}, \bibinfo{author}{Morandotti, R.},
  \bibinfo{author}{Gaeta, A.~L.} \& \bibinfo{author}{Lipson, M.}
\newblock \bibinfo{title}{New {CMOS}-compatible platforms based on silicon
  nitride and {Hydex} for nonlinear optics}.
\newblock \emph{\bibinfo{journal}{Nature Photonics}}
  \textbf{\bibinfo{volume}{7}}, \bibinfo{pages}{597--607}
  (\bibinfo{year}{2013}).
\newblock \urlprefix\url{https://www.nature.com/articles/nphoton.2013.183}.

\bibitem{ramelow_silicon-nitride_2015}
\bibinfo{author}{Ramelow, S.} \emph{et~al.}
\newblock \bibinfo{title}{Silicon-{Nitride} {Platform} for {Narrowband}
  {Entangled} {Photon} {Generation}}.
\newblock \emph{\bibinfo{journal}{arXiv:1508.04358 [physics,
  physics:quant-ph]}}  (\bibinfo{year}{2015}).
\newblock \urlprefix\url{http://arxiv.org/abs/1508.04358}.
\newblock \bibinfo{note}{ArXiv: 1508.04358}.

\bibitem{guo_parametric_2017}
\bibinfo{author}{Guo, X.} \emph{et~al.}
\newblock \bibinfo{title}{Parametric down-conversion photon-pair source on a
  nanophotonic chip}.
\newblock \emph{\bibinfo{journal}{Light: Science \& Applications}}
  \textbf{\bibinfo{volume}{6}}, \bibinfo{pages}{e16249--e16249}
  (\bibinfo{year}{2017}).
\newblock \urlprefix\url{https://www.nature.com/articles/lsa2016249}.

\bibitem{ma_silicon_2017}
\bibinfo{author}{Ma, C.} \emph{et~al.}
\newblock \bibinfo{title}{Silicon photonic entangled photon-pair and heralded
  single photon generation with {CAR} $>$ 12,000 and g$^{\textrm{(2)}}$(0)
  $<$ 0.006}.
\newblock \emph{\bibinfo{journal}{Optics Express}}
  \textbf{\bibinfo{volume}{25}}, \bibinfo{pages}{32995--33006}
  (\bibinfo{year}{2017}).
\newblock
  \urlprefix\url{https://www.osapublishing.org/oe/abstract.cfm?uri=oe-25-26-32995}.
\newblock \bibinfo{note}{Publisher: Optical Society of America}.

\bibitem{steiner_ultrabright_2021}
\bibinfo{author}{Steiner, T.~J.} \emph{et~al.}
\newblock \bibinfo{title}{Ultrabright entangled-photon-pair generation from an
  al ga as -on-insulator microring resonator}.
\newblock \emph{\bibinfo{journal}{{PRX} Quantum}} \textbf{\bibinfo{volume}{2}},
  \bibinfo{pages}{010337} (\bibinfo{year}{2021}).
\newblock \urlprefix\url{https://link.aps.org/doi/10.1103/PRXQuantum.2.010337}.

\bibitem{luo_direct_2015}
\bibinfo{author}{Luo, K.-H.} \emph{et~al.}
\newblock \bibinfo{title}{Direct generation of genuine single-longitudinal-mode
  narrowband photon pairs}.
\newblock \emph{\bibinfo{journal}{New Journal of Physics}}
  \textbf{\bibinfo{volume}{17}}, \bibinfo{pages}{073039}
  (\bibinfo{year}{2015}).
\newblock \urlprefix\url{https://doi.org/10.1088/1367-2630/17/7/073039}.
\newblock \bibinfo{note}{Publisher: IOP Publishing}.

\bibitem{luo_nonlinear_nodate}
\bibinfo{author}{Luo, K.-H.} \emph{et~al.}
\newblock \bibinfo{title}{Nonlinear integrated quantum electro-optic circuits}.
\newblock \emph{\bibinfo{journal}{Science Advances}}
  \textbf{\bibinfo{volume}{5}}, \bibinfo{pages}{eaat1451}
  (\bibinfo{year}{2019}).
\newblock \urlprefix\url{https://www.science.org/doi/10.1126/sciadv.aat1451}.
\newblock \bibinfo{note}{Publisher: American Association for the Advancement of
  Science}.

\bibitem{Davanco2017}
\bibinfo{author}{Davanco, M.} \emph{et~al.}
\newblock \bibinfo{title}{Heterogeneous integration for on-chip quantum
  photonic circuits with single quantum dot devices}.
\newblock \emph{\bibinfo{journal}{Nature Communications}}
  \textbf{\bibinfo{volume}{8}}, \bibinfo{pages}{889} (\bibinfo{year}{2017}).
\newblock \urlprefix\url{https://doi.org/10.1038/s41467-017-00987-6}.

\bibitem{sun_adiabaticity_2009}
\bibinfo{author}{Sun, X.}, \bibinfo{author}{Liu, H.-C.} \&
  \bibinfo{author}{Yariv, A.}
\newblock \bibinfo{title}{Adiabaticity criterion and the shortest adiabatic
  mode transformer in a coupled-waveguide system}.
\newblock \emph{\bibinfo{journal}{Opt. Lett., {OL}}}
  \textbf{\bibinfo{volume}{34}}, \bibinfo{pages}{280--282}
  (\bibinfo{year}{2010}).
\newblock
  \urlprefix\url{https://www.osapublishing.org/abstract.cfm?uri=ol-34-3-280}.

\bibitem{Ulhaq2013}
\bibinfo{author}{Ulhaq, A.} \emph{et~al.}
\newblock \bibinfo{title}{Detuning-dependent mollow triplet of a
  coherently-driven single quantum dot}.
\newblock \emph{\bibinfo{journal}{Optics Express}}
  \textbf{\bibinfo{volume}{21}}, \bibinfo{pages}{4382--4395}
  (\bibinfo{year}{2013}).
\newblock
  \urlprefix\url{http://www.opticsexpress.org/abstract.cfm?URI=oe-21-4-4382}.

\bibitem{konthasinghe_coherent_2012}
\bibinfo{author}{Konthasinghe, K.} \emph{et~al.}
\newblock \bibinfo{title}{Coherent versus incoherent light scattering from a
  quantum dot}.
\newblock \emph{\bibinfo{journal}{Physical Review B}}
  \textbf{\bibinfo{volume}{85}}, \bibinfo{pages}{235315}
  (\bibinfo{year}{2012}).
\newblock \urlprefix\url{https://link.aps.org/doi/10.1103/PhysRevB.85.235315}.
\newblock \bibinfo{note}{Publisher: American Physical Society}.

\bibitem{nguyen_optically_2012}
\bibinfo{author}{Nguyen, H.~S.} \emph{et~al.}
\newblock \bibinfo{title}{Optically gated resonant emission of single quantum
  dots}.
\newblock \emph{\bibinfo{journal}{Physical Review Letters}}
  \textbf{\bibinfo{volume}{108}}, \bibinfo{pages}{057401}
  (\bibinfo{year}{2012}).

\bibitem{proux_measuring_2015}
\bibinfo{author}{Proux, R.} \emph{et~al.}
\newblock \bibinfo{title}{Measuring the {Photon} {Coalescence} {Time} {Window}
  in the {Continuous}-{Wave} {Regime} for {Resonantly} {Driven} {Semiconductor}
  {Quantum} {Dots}}.
\newblock \emph{\bibinfo{journal}{Physical Review Letters}}
  \textbf{\bibinfo{volume}{114}}, \bibinfo{pages}{067401}
  (\bibinfo{year}{2015}).
\newblock
  \urlprefix\url{https://link.aps.org/doi/10.1103/PhysRevLett.114.067401}.

\bibitem{muller}
\bibinfo{author}{Muller, A.} \emph{et~al.}
\newblock \bibinfo{title}{Resonance fluorescence from a coherently driven
  semiconductor quantum dot in a cavity}.
\newblock \emph{\bibinfo{journal}{Physical Review Letters}}
  \textbf{\bibinfo{volume}{99}}, \bibinfo{pages}{187402}
  (\bibinfo{year}{2007}).
\newblock
  \urlprefix\url{https://link.aps.org/doi/10.1103/PhysRevLett.99.187402}.

\end{thebibliography}

 \end{document}